\documentclass[a4paper, 11pt]{article}
\usepackage[utf8]{inputenc}
\usepackage{jcappub}
\usepackage{graphicx}  % needed for figures
\usepackage{dcolumn}   % needed for some tables
\usepackage{bm}        % for math
\usepackage{amssymb}
\usepackage{amsmath}   % for math
\usepackage{hyperref}

\hypersetup{colorlinks,linkcolor={blue},citecolor={blue},urlcolor={blue}}
\usepackage{float}

\newcommand{\Mpl}{M_{_{\rm Pl}}}

\def\n{\nabla}

\def\beq{\begin{equation}}
	\def\eeq{\end{equation}}
\def\bea{\begin{eqnarray}}
	\def\eea{\end{eqnarray}}

\newcommand{\viz}{\textit{viz.~}}

	\title{Unifying inflationary and reheating solution}% Force line breaks with \\
	\author{Manjeet Kaur, Debottam Nandi, Sharath Raghavan B}
	\emailAdd{mkaur1@physics.du.ac.in}
	
	\emailAdd{dnandi@physics.du.ac.in}
	
	\emailAdd{sharathkuttikol@gmail.com\\
	Current address: Pranavam house, Kuttikol post, Kasargod, Kerala, India, 671541}

	\affiliation{Department of Physics and Astrophysics, University of Delhi, Delhi 110007, India}
	
	\abstract{
		The conventional background solution for the evolution of a single canonical inflaton field performs admirably in extreme scenarios such as the slow-roll phase (where the slow-roll parameter is much less than one) and the deep reheating era (where the Hubble parameter is much smaller than the effective mass of the potential and the field oscillates around the minimum of the potential), but fails to accurately depict the dynamics of the Universe around the end of inflation and the initial oscillatory phases. This article proposes a single, unified, model-independent, parametrized analytical solution for such models that bridges the gap between these two extremes, providing a near-accurate comprehensive description of the evolution of the Universe. This novel strategy has the potential to substantially enhance both quantitative and qualitative cosmological observational predictions, and, as a consequence, can further constrain the inflationary models more effectively using future observations.}
	
\begin{document}
	\maketitle
	%%%%%%%%%%%%%%%%%%%%%%%%%%%%%%%%%%%%%%%%%%%
	%%%%%%%%%%%%%%%%%%%%%%%%%%%%%%%%%%%%%%%%%%%
	%%%%%%%%%%%%%%%%%%%%%%%%%%%%%%%%%%%%%%%%%%%
	\section{Introduction}
	%{
		% \begin{itemize}
			%     \item The method shows systematically how to arise at different levels of Hubble parameter values using $\theta$, i.e., $\theta = (2n + 1)\pi/2, n \in N$ indicates the field is at the bottom of the potential, whereas $\theta = 2n\pi$ denotes the field reaches the highest level of the potential during reheating.
			%     \item In quantitative analysis of preheating, it also helps improving the accuracy of the $n_{\rm{s}}$ vs. $T_{\rm{re}}$ relation.
			%     \item In qualitative analysis, it helps to incorporate the effect of the end of inflation and the beginning of the reheating phase.
			%     \item It may also lead to indicate the PBH production in different scenarios at the end of inflation.
			%     \end{itemize}}

	%The standard Big Bang theory of the Universe, although consistent with general relativity and in line with the observed Hubble expansion, has been deemed insufficient in explaining certain puzzles pertaining to the early Universe, such as the horizon problem and the flatness problem.
	
	The inflationary paradigm \cite{STAROBINSKY198099, Guth:1981, Sato:1981, Mukhanov:1981xt, LINDE1982389, HAWKING1982295, STAROBINSKY1982175, Guth:1982, Albrecht-Steinhardt:1982, Albrecht:1982mp, Linde:1983gd, VILENKIN1983527, Bardeen:1983, 1990-Kolb.Turner-Book, Mukhanov:1990me, Liddle:1994dx, 2005hep.th....3203L, Bassett:2005xm, Sriramkumar:2009kg, Baumann:2009ds, Linde:2014nna, Martin:2015dha, Ade:2015lrj, Ade:2015ava}, a brief period of accelerated expansion during the early Universe, not only overcomes the early Universe puzzles, such as the horizon and the flatness problems, but also explains the observational constraints \cite{Planck:2018jri, BICEP:2021xfz, Galloni:2022mok}. Within the paradigm, the single canonical scalar field-driven slow-roll inflationary models are the most successful ones, where the inflaton (scalar) field slowly rolls down its potential, resulting in a quasi-exponential expansion of the Universe.
	As the field approaches the bottom of the potential, the inflationary stage ends, and the field smoothly starts oscillating around the minimum of its effective potential. At this stage, it couples to other (standard) particles and the inflaton field decays into those elementary particles, resulting in the transfer of energy from the inflaton field to those particles. This era is referred to as the reheating epoch \cite{mclachlan1947theory, Albrecht:1982mp, Dolgov:1982th, Abbott:1982hn, Traschen:1990sw, Kofman:1994rk, Shtanov:1994ce, Kofman:1997yn, Liddle:2003as, Bassett:2005xm, Martin:2006rs, Lorenz:2007ze, Korsch2008, Allahverdi:2010xz, Martin:2010kz, Adshead:2010mc, Mielczarek:2010ag, Easther:2011yq, Choudhury:2013qza, Amin:2014eta, Dai:2014jja, Martin:2014nya, Domcke:2015iaa, Maity:2016uyn, Lozanov:2016hid, Maity:2018qhi, Kabir:2016kdh, Maity:2019ltu, Odintsov:2023lbb,Chowdhury:2023jft}. When the inflaton field completely decays, the reheating era ends, and the Universe enters into the known radiation-dominated era \cite{Albrecht:1982mp, Abbott:1982hn, Kofman:1994rk, ElBourakadi:2021blc}.
	% The aforementioned particles engage in mutual interactions, leading to their eventual achievement of thermal equilibrium at a specific temperature denoted as $T$. The aforementioned process reaches its ultimate stage when nearly all, if not all, of the energy possessed by the classical inflaton field is transferred to the thermal energy of elementary particles. The temperature of the Universe during this particular phase is commonly referred to as the reheating temperature, denoted as $T_{\rm{re}}$, which denotes the onset of the radiation-dominated epoch.
	
	For slow-roll inflation to occur, the magnitude of the slow-roll parameters $\epsilon_{\rm{1}}$ and $\epsilon_{\rm{2}}$ must be extremely close to zero. These conditions simply refer to the exceedingly slowly changing (decreasing) Hubble parameter. This, in turn, leads to the quasi-exponential expansion of the Universe. To better comprehend the dynamics, consider the chaotic inflationary model with the potential \cite{Linde:1983gd}: 
	
	$$V(\phi) = \frac{1}{2} m^2 \phi^2,$$
	where $m$ is the mass of the inflaton field $\phi$. The slow-roll condition is met for $|\phi| \gg 1$, and at this stage, the evolution of the Hubble parameter as a function of cosmic time $t$ can be expressed as $H(t) = H_{0} -  \frac{1}{3} m^2 t$, where $H_{0}$ is a constant (as will be demonstrated later). It results in the scale factor solution $a(t) \propto \exp{(H_{0} t  - \frac{1}{6} m^2 t^2)}$, i.e., a solution close to exponential. As is evident, during the slow-roll era, $H \gg m$. The inflationary era ends when $\epsilon_{\rm{1}} = 1$, which implies $H \sim m$. Shortly after the end of inflation, the field oscillates, and the reheating epoch begins. During the reheating period, the scalar field $\phi$ oscillates around the minimum, $\phi = 0$, and as a consequence, the Hubble parameter decreases as $H(t) \sim 2/(3t)$, indicating $H \ll m$.
	
	These characteristics can be generalized to any inflationary potential: during the slow-roll epoch, the magnitude of the Hubble parameter is significantly higher  than the effective mass of the field, i.e., second derivative of the potential. The analytical solution, in this era, i.e., the slow-roll solution, is well-known. Inflation ends approximately when the Hubble parameter equals the mass and, during reheating, the Hubble parameter falls significantly over the mass and decreases very quickly with time. In this period also, we can represent the approximate analytical solution of the reheating epoch. However, both the slow-roll inflationary solution as well as standard reheating solution fail to address the smooth transition from the slow-roll regime to the oscillations, and thus, the dynamics, as well as the complexities such as the study of the perturbations at the end of inflation and the beginning of the reheating process, are incomplete. This is because the slow-roll approximation fails near the end of inflation, and hence, the slow-roll solution does not justify the genuine solution near the end of inflation. Similarly, in solving the dynamics in the reheating epoch, we use the approximation that the Hubble parameter is subdominant over the effective mass of the potential (e.g., $H \ll m$ in the case of chaotic inflation). Thus, the reheating solution only accounts for the asymptotic oscillatory solution, while the solution near the first and subsequent oscillations misses the true dynamics.
	
	Such difficulties also affect the inflationary and reheating constraints. As we know, there are primarily two ways to investigate the reheating era: quantitative \cite{Martin:2006rs, Lorenz:2007ze, Martin:2010kz, Adshead:2010mc, Mielczarek:2010ag, Easther:2011yq, Dai:2014jja, Domcke:2015iaa, Lozanov:2016hid, Kabir:2016kdh, Liddle:2003as, Martin:2014nya, Maity:2016uyn, Maity:2018qhi, Maity:2019ltu, Nandi:2019xve} and qualitative analysis \cite{Albrecht:1982mp, Dolgov:1982th, Abbott:1982hn, Traschen:1990sw, Kofman:1994rk, Kofman:1997yn, Shtanov:1994ce, Bassett:2005xm, Allahverdi:2010xz, Amin:2014eta, mclachlan1947theory, Korsch2008, Maity:2018qhi}. 
	 Quantitative analysis operates under the assumption that the Universe behaves effectively during the reheating stage and its objective is to constrain the effective equation of state $w_{\rm re}$ and the duration of reheating $N_{\rm re}$ (or the temperature $T_{\rm re}$ at the end of reheating) by utilizing perturbations generated during the inflationary regime \cite{Turner:1983he, Martin:2013tda, Dai:2014jja, Martin:2014nya}. In short, in this study, the late-time dynamics (after the end of inflation) determine the constraint on the parameter $N_k$, which represents the time interval between the end of inflation and the epoch at which the pivot scale leaves the Hubble horizon.  Due to the fact that inflationary observables are dependent on $N_k$, even a slight modification to the dynamics resulting from the aforementioned accuracy will have an immediate impact on the theoretical prediction of inflationary observables. Indeed, subsequent sections will demonstrate that even increasing the precision of $\mathcal{O}(1)$ for $N_k$ can lead to an enhancement in the spectral index $\Delta n_s \sim 10^{-3}$. As a result, models can be rigorously constrained by using the forthcoming experiments such as PRISM \cite{PRISM:2013ybg} and EUCLID \cite{EuclidTheoryWorkingGroup:2012gxx}, cosmic 21-cm surveys \cite{Mao:2008ug}, and CORE experiments \cite{CORE:2016ymi}, which offer a precision enhancement of $10^{-3}$ in $n_s$. 
		
	On the contrary, in qualitative analysis, we approximate the background evolution during the reheating (oscillatory) phase and thereafter examine the decay of the inflaton field into other elementary particles. Parametric resonance, as described by the Mathieu equation, is the process responsible for the generation of these particles. This phenomenon accurately explains the microscopic dynamics of the reheating epoch. As previously mentioned in the quantitative analysis, obtaining an exact solution for the background, rather than an approximate one, can modify the Mathieu equation and therefore, the solution can greatly enhance our understanding of the microphysics of this time period. To summarize, the accurate dynamics can significantly improve the physical prediction through both the qualitative and quantitative study of reheating, when compared to the previous methods that relied on slow-roll and reheating (oscillatory) phase approximations.

	This paper thus focuses on the analytical unification of the slow-roll inflationary and the reheating solutions into a single unified solution using trivial parametrization method, which eventually captures the true dynamics not only during the slow-roll and asymptotic reheating epoch but also during the intermediate junction between the two epochs, i.e., during the end of inflation and the beginning of reheating epochs. The suggested method provided in this article is also model-independent, meaning that it can account for both small and large field models. Although the method can be applied for a large number of models, we must emphasize that, at this moment, it is difficult to employ it for models with non-trivial features such as the models that can produce primordial black holes. Nonetheless, it is still a useful method for constraining other existing simpler models. While full numerical simulations for models with a single parameter are possible, studying multi-parameter models presents a significant challenge due to their high numerical complexity. In that case, our proposed method can also be easily implemented. As we will demonstrate in following sections, we must now acknowledge that our suggested approach is also not an exact solution but rather a close approximation that near-accurately depicts the dynamics. Nevertheless, assuming the  near-accurate unified analytical solution is achieved, then, as previously mentioned, it can offer a more comprehensive view of the unified early Universe solution and can also yield better constraints on parameters through the use of both qualitative and quantitative methods of reheating. It can even be studied in special cases of producing primordial black holes (PBHs), primordial gravitational waves (PGWs), and other scenarios where using parametric resonance during reheating; perturbed modes can be enhanced \cite{Jedamzik:2010dq, Sharma:2018kgs, Martin:2019nuw, Auclair:2020csm, Haque:2020bip, Bamba:2021wyx}. It is crucial to acknowledge that the work in this article only focuses on the background dynamics of the Universe, and the full analysis of the perturbations is reserved for future work as it is beyond the scope of this work.

	For such analysis, in this article, we consider the single canonical scalar field minimally coupled to gravity and provide a single analytical solution for the early Universe. In doing so, rather than working in the phase space consisting of  $\{\phi, \dot{\phi}\}$, we conveniently choose the parameter  $\theta$, which represents the phase of the oscillatory solution, and the Hubble parameter $H$. Since the particle production (resonance) occurs at the minimum of the potential and it is difficult to be certain of the same using cosmic time $t$, the introduction of the coordinate $\theta$ mitigates this issue. This method systematically shows how to arise at different levels of Hubble parameter values using $\theta$, i.e., $\theta = (2n + 1)\pi/2, n \in N$ indicates the field is at the bottom of the potential, whereas $\theta = n\pi$, $n \in N$ denotes the field reaches the highest level of the potential, where the field velocity $\dot{\phi}$ vanishes during reheating. Knowing this instances helps in solving the system to achieve a detailed picture, as we will show in later sections.
	
	The following is how the article is written. The action responsible for the early Universe dynamics is defined in Sec. \ref{sec:gen}, and the generic background equations are provided. Sec. \ref{sec:scalar field in diff reg} demonstrates how to obtain the usual slow-roll and reheating solutions. In this part, we also present the phase space, which comprises of $\theta$ and $H$, as well as the asymptotic reheating solution, which is well-known. In Sec. \ref{sec:chaotic-slow-roll}, we extend the phase space solution, which is used to get the reheating oscillatory solution, to the slow-roll phase, and later we present our primary work in the next Sec. \ref{sec:full-chaotic}. In this section,  we unify the inflationary and reheating solutions, i.e., the complete, yet model-independent, solution of the homogeneous Universe dominated by the canonical inflaton field, and demonstrate the result for the chaotic inflationary model using a simple yet straightforward method. We demonstrate that different values of $\theta$ represent different instances of the early Universe. 
	% For example, $\theta = 0$ indicates that inflation has begun, $\theta = \sin^{-1}(1/\sqrt{3})$ indicates that the inflation has ended, $\theta = (2n + 1) \pi/2$ indicates that the field is at the bottom of the potential, and so on. 
	Thus, by providing the solution of $H$ (and other background variables) in terms of $\theta$, we explicitly imply that, in each of the instances listed, we know the value of the $H$ (and other background variables), which clearly aids in understanding the dynamics. We also present the solution of $\theta$ in terms of cosmic time $t$, completing the solution. In Sec. \ref{sec:gen sol}, we extend our result and investigate several inflationary models, demonstrating that our method brilliantly provides the entire background solution of the dynamics, i.e., from slow-roll to reheating solution with a smooth transition, and we study the observational consequences in Sec. \ref{sec:obs}. Finally, in Sec. \ref{sec:conclu}, we conclude our work.		

	A few words about our conventions and notations are in order at this stage of our discussion. In this work, we work with the natural units such that $\hbar = c = 1$, and we define the reduced Planck mass to be $\Mpl \equiv (8\pi G)^{-1/2} = 1$. We adopt the metric signature of $(-, +, +, +)$. Also, we should mention that, while the Greek indices are contracted with the metric tensor $g_{\mu \nu}$, the Latin indices are contracted with the Kronecker delta $\delta_{i j}.$ Moreover, we shall denote the partial and the covariant derivatives as $\partial$ and $\n$. The overdots and overprimes denote derivatives with respect to the cosmic time $t$ and the conformal time $\eta$ associated with the Friedmann-Lema\^{\i}tre-Robertson-Walker (FLRW) line-element, respectively.
	
	%%%%%%%%%%%%%%%%%%%%%%%%%%%%%%%%%%%%%%%%%%%
	%%%%%%%%%%%%%%%%%%%%%%%%%%%%%%%%%%%%%%%%%%%
	%%%%%%%%%%%%%%%%%%%%%%%%%%%%%%%%%%%%%%%%%%%
	\section{General Equations}\label{sec:gen}
	
	Let us first consider a single canonical scalar field $\phi$ minimally coupled to the gravity with a potential $V(\phi)$, specified by the action
	\begin{eqnarray}\label{eq:gen-can-action}
		S=\frac{1}{2}\int d^4 x \sqrt{-g} ~\left(R - g^{\mu\nu}\partial_{\mu}\phi ~ \partial_{\nu}\phi-2V(\phi)\right),
	\end{eqnarray}
	where $R$ is the Ricci scalar. The corresponding equations of motion, i.e., Einstein's equations and the equation of the scalar field, can be written as
	\begin{eqnarray}\label{eq:eins-eq}
		R_{\mu\nu}-\frac{1}{2}g_{\mu\nu} R &=& T_{\mu\nu (\phi)},\\
		\label{eq:cont-eq}
		\nabla_\mu T^{\mu\nu}_{{\rm{(\phi)}}}&=&0,
	\end{eqnarray}
	where $T^\mu_{\,\nu{\rm{(\phi)}}}$ is the stress-energy tensor corresponding to the $\phi$ field:
	\begin{equation}\label{eq:EM-tensor}
		T_{\mu\nu{\rm{(\phi)}}}=  \partial_\mu\phi\ \partial_\nu \phi-g_{\mu\nu}\left(\frac{1}{2}\partial_\lambda\phi \ \partial^\lambda \phi+V(\phi)\right).
	\end{equation}
	%The equation of motion for the scalar field is simply
	% \begin{eqnarray}
		% %R_{\mu\nu}- \frac{1}{2}g_{\mu\nu } R&=&\left(\partial_{\mu}\phi \partial_{\nu}\phi- \frac{1}{2}g_{\mu\nu}\partial^{\lambda}\phi\partial_{\lambda}\phi\right)-g_{\mu\nu}V,\nonumber\\
		% %&&\\
		% \Box \phi -V_{,\phi} &=& 0,
		% \end{eqnarray}
	
	%$R_{\mu\nu}$is the Ricci tensor,~$R\equiv g^{\mu\nu}R_{\mu\nu}$ is the Ricci scalar.
	\noindent Using the FLRW line element, describing the homogeneous and isotropic Universe in cosmic time $t$:
	\begin{eqnarray}\label{eq:FLRW_metric}
		ds^2=-{\rm d}t^2+ a^2(t){\rm d\bf x}^2,
	\end{eqnarray}
	where $a(t)$ is the scale factor,  Eqs. \eqref{eq:eins-eq} and \eqref{eq:cont-eq} can be reduced to the following forms:
	\begin{eqnarray}
		\label{eq:energy-eq}
		&&3 H^2 = \frac{1}{2} \dot {\phi}^2 + V(\phi),\\
		\label{eq:acc-eq}
		&&\dot{H} = -\frac{1}{2}\dot{\phi}^2, \\
		\label{eq:phi}
		&&\ddot{\phi} + 3 H \dot{\phi} + V_{,\phi} = 0.
	\end{eqnarray}
	where, $H \equiv\dot{a}/a$ is the Hubble parameter and $A_{,\phi} \equiv \partial A/ \partial{\phi}$. As one can see, the first one is a constrained equation, and between the other two, one of them is independent, leaving the degrees of the freedom of the system to one with a single evolutionary equation:
	\begin{eqnarray}\label{eq:single-phi-eq}
		\ddot{\phi} + \sqrt{\frac{3}{2}}\dot{\phi} \sqrt{\dot{\phi}^2 + 2 V} + V_{,\phi} = 0.
	\end{eqnarray}
The above equation is highly nonlinear; therefore, obtaining its general solution is exceedingly challenging. Using certain approximations, Eq. \eqref{eq:single-phi-eq} can be solved under various conditions, as demonstrated in the following section. The primary objective of this article is,  contrary to the conventional method of solving in various epochs (or conditions), to provide a complete solution of the above equation for a variety of models, as will be demonstrated later.
	
	Finally, we now can define the two slow-roll parameters $\epsilon_{\rm{1}}$ and $\epsilon_{\rm{2}}$ as
	\begin{eqnarray}
		\label{eq:slow_roll_params-def}
		\epsilon_{\rm{1}} \equiv - \frac{\dot{H}}{H^2},\qquad 
		\epsilon_{\rm{2}} \equiv \frac{\dot{\epsilon_{\rm{1}}}}{H \epsilon_{\rm{1}}}.
	\end{eqnarray}
	These slow-roll parameters play a crucial role in defining the dynamics in the early Universe, mainly during slow-roll inflationary evolution. In the next section, with the help of these parameters, we will establish the inflationary as well as the reheating dynamics.
	
	%%%%%%%%%%%%%%%%%%%%%%%%%%%%%%%%%%%%%%%%%%%
	%%%%%%%%%%%%%%%%%%%%%%%%%%%%%%%%%%%%%%%%%%%
	%%%%%%%%%%%%%%%%%%%%%%%%%%%%%%%%%%%%%%%%%%%
	\section{Scalar field solutions in different regimes}\label{sec:scalar field in diff reg}
	
	% As the general background equations are now known and already given in the previous section, hence, if we provide specific potential as well as the initial conditions, using those equations, one can obtain the evolution of the Universe.  
	Given that the generic background equations are now known and have already been presented in the preceding section, one can obtain the evolution of the Universe using those equations for a given potential as well as the initial conditions. Let us first discuss the slow-roll inflation.
	
	%%%%%%%%%%%%%%%%%%%%%%%%%%%%%%%%%%%%%%%%%%%
	%%%%%%%%%%%%%%%%%%%%%%%%%%%%%%%%%%%%%%%%%%%
	%%%%%%%%%%%%%%%%%%%%%%%%%%%%%%%%%%%%%%%%%%%
	\subsection{Slow-roll Equations}
	In order to achieve a slow-roll inflation in the early Universe, the above-mentioned slow-roll parameters have to be extremely small, i.e.,
	\begin{eqnarray}\label{eq:slow-roll-cond}
		&& \epsilon_{\rm{1}}\ll1, \qquad\qquad \epsilon_{\rm{2}}\ll 1.
	\end{eqnarray}
	The first condition in the above equation leads to $\dot{\phi}^2 \ll H^2,$ meaning the field velocity is small compared to the potential, thus the name slow-roll and the second condition leads to $\ddot{\phi}\ll \dot{\phi}H $, implying that the field acceleration is extremely small, i.e., the first condition stay relevant for a sufficient time, which constrains the Eqs. \eqref{eq:energy-eq} and \eqref{eq:phi} to,
	\begin{eqnarray}
		\label{eq:slow-roll-eqs} 
		3H^2 \simeq V(\phi),\qquad\qquad
		3H \dot{\phi} \simeq -V_{,\phi}.
	\end{eqnarray}
	These equations define the dynamics corresponding to the slowly rolling scalar fields. The two slow-roll conditions can then also be expressed directly in terms of the shape of inflationary potential as
	\begin{eqnarray}
		\label{eq:slow_roll_in_V}
		\epsilon_{\rm{1}} \simeq \frac{1}{2}\left(\frac{V_{,\phi}}{V}\right)^2\qquad
		\epsilon_{\rm{2}} \simeq 2\left(\frac{V_{,\phi}^2}{V^2}-\frac{V_{,\phi\phi}}{V}\right),
	\end{eqnarray}
	where, $A_{,xx}=\frac{\partial^2 A}{\partial x^2}$. Given the potential as well as the field value, if the aforementioned slow-roll parameters satisfy the slow-roll conditions \eqref{eq:slow-roll-cond}, then one can ensure the Universe is in the slow-roll stage, and the specific dynamics can be obtained by solving the slow-roll equations \eqref{eq:slow-roll-eqs}.
	
	%%%%%%%%%%%%%%%%%%%%%%%%%%%%%%%%%%%%%%%%%%%
	%%%%%%%%%%%%%%%%%%%%%%%%%%%%%%%%%%%%%%%%%%%
	%%%%%%%%%%%%%%%%%%%%%%%%%%%%%%%%%%%%%%%%%%%
	To illustrate the slow-roll inflationary scenario, consider the simplest model with quadratic potential, i.e., the chaotic inflation model:
	\begin{eqnarray}
		V(\phi)= \frac{1}{2}m^2 \phi^2,
	\end{eqnarray} 
	where, $m$ is the mass of the scalar field $\phi$. Using Eqs. \eqref{eq:slow_roll_in_V}, the slow-roll parameters are:
	\begin{eqnarray}{\label{eq:epsilon_slow_roll}}
		\epsilon_{\rm{1}} \simeq \frac{2}{\phi^2}, \qquad\qquad \epsilon_{\rm{2}} \simeq \frac{4}{\phi^2},
	\end{eqnarray}
	which implies that, only when $|\phi| \gg 1,$ the slow-roll conditions are met. Only in this regime  the slow-roll equations \eqref{eq:slow-roll-eqs} can be used to obtain the dynamics, and they are given as
	\begin{eqnarray}
		\label{eq:slow_roll_friedman_chaotic}
		3H^2\simeq \frac{1}{2}m^2\phi^2,\qquad\qquad\dot{\phi}\simeq-m \sqrt{\frac{2}{3}}.
	\end{eqnarray}
	As slow-roll can be achieved only during $|\phi| \gg 1,$ during this regime, $H \gg m.$ As a result, the solution to these equations, i.e., the slow-roll solutions, can be obtained as
	\begin{eqnarray}
		\label{eq:slow_roll_chaotic}
		&&\phi\simeq\phi_{i}-\sqrt{\frac{2}{3}}m t,\\
		\label{eq:slow_roll_chaotic_H}
		&&H\simeq\left(\frac{1}{\sqrt{6}} m \phi_{i}-\frac{1}{3}m^2 t\right),
	\end{eqnarray}
	and the solution of the scale factor during slow-roll can now be expressed as
	\begin{equation}
		a(t) \simeq a_{i}\exp{\left(\frac{1}{\sqrt{6}} m \phi_{i}t-\frac{1}{6}m^2 t^2\right)}.
	\end{equation}
	$\phi_{i}$ and $a_{i}$ are the initial values of $\phi$ and $a$ at $t=0$. Note that the 
	% \begin{figure}[H]
		% \includegraphics[scale=0.4]{Plots/ChaoticNum/phinum.pdf}% Here is how to import EPS art
		% \caption{\label{fig:chaotic_phinum} $\phi$ behavior during inflation for a initial value of $\phi$, $\phi_{i}=15$ .}
		% \end{figure}
	% \begin{figure}[H]
		% \includegraphics[scale=0.4]{Plots/ChaoticNum/Hnum.pdf}% Here is how to import EPS art
		% \caption{\label{fig:chaotic_Hnum} $H$ behavior during inflation for $\phi_{i}=15$ .}
		% \end{figure}
	inflation ends at $\epsilon_{\rm{1}}=1$, and by assuming the slow-roll dynamics holds till the end of inflation, with the help of Eq. \eqref{eq:epsilon_slow_roll}, one can find the field value at the end of inflation, and it is $|\phi_{\rm{end}}|\simeq \sqrt{2}$. In that case, we can also solve for cosmic time $t_{\rm{end}}$ denoting the end of inflation as
	\begin{eqnarray}
		t_{\rm{end}}\simeq \sqrt{\frac{3}{2}}\ \frac{\phi_{i}}{m}
	\end{eqnarray}
	where $t=t_{\rm{end}}$ corresponds to the end of slow-roll inflation, and we assume $|\phi_{i}| \gg |\phi_{\rm{end}}|$ which is required for slow-roll inflation. The Hubble parameter, then, at the end of inflation, $H_{\rm{end}}$, is
	\begin{equation}
		H_{\rm{end}} \simeq \frac{1}{\sqrt{3}} m.
	\end{equation}
	
	Please note that very close to the end of inflation, i.e., $|\phi| = \sqrt{2},$ the slow-roll parameters do not obey the slow-roll condition as  $\epsilon_{\rm{1}}, \epsilon_{\rm{2}} \sim 1.$ In fact, $\epsilon_{\rm{2}}$ becomes one at $|\phi| \simeq 2,$ even before the end of inflation. Therefore, these solutions do not represent the true solutions at the end of inflation and thereafter.
	% Now the slow-roll parameter \eqref{eq:first_slow_roll_in_V} takes the form,
	% \begin{eqnarray}
		%\epsilon_{\rm{1}}&=&\frac{2}{\left(-\sqrt{\frac{2}{3}}m t+\phi_{i}\right)^2}.
		% \end{eqnarray}
	
	% \begin{figure}[H]
		% \includegraphics[scale=0.4]{Plots/ChaoticNum/epsnum.pdf}% Here is how to import EPS art
		% \caption{\label{fig:chaotic_epsnum} $\epsilon_{\rm{1}}$ behavior during inflation for a initial value of $\phi$, $\phi_{i}=15$ .}
		% \end{figure}
	% Now at the end of inflation is when $\epsilon=1$,
	% \begin{eqnarray}
		% \left(-\sqrt{\frac{2}{3}}m t+\phi_{i}\right)^2=2,
		% \end{eqnarray}
	
	%%%%%%%%%%%%%%%%%%%%%%%%%%%%%%%%%%%%%%%%%%%
	%%%%%%%%%%%%%%%%%%%%%%%%%%%%%%%%%%%%%%%%%%%
	%%%%%%%%%%%%%%%%%%%%%%%%%%%%%%%%%%%%%%%%%%%
	\subsection{Reheating}
	Deep within the slow-roll regime, the first slow-roll parameter is very close to zero, i.e., $\epsilon_{\rm{1}} \ll 1$, by definition. Nonetheless, as the field value decreases, $\epsilon_{\rm{1}}$ and $\epsilon_{\rm{2}}$ increase (see, for instance, Eq. \eqref{eq:epsilon_slow_roll}), and inflation ceases when $\epsilon_{\rm{1}}$ equals 1. Consequently, during the slow-roll epoch, the potential energy predominates over the kinetic energy, and as the inflation approaches the end of it, the contribution to the kinetic energy increases while the contribution to the potential energy decreases until they are almost equal at the end. The field then begins to oscillate around the minimum of the potential, and the reheating phase commences. To derive an analytical solution for this regime, it is easier to work with the phase space orientation of the field, i.e., $\theta$ and $H$, as opposed to $\phi$ and $\dot{\phi}$ \cite{Mukhanov:1990me, Lozanov:2019jxc, Kwapisz:2019cxq, Mukhanov:2005sc, Urena-Lopez:2016yon}. To illustrate this, let us define
	\begin{eqnarray}
		\label{eq:phi_and_phidot}
		\frac{\dot{\phi}}{\sqrt{6 }}\equiv -H \sin{\theta}, 
		\qquad\qquad \sqrt\frac{V}{3}\equiv H\cos{\theta},
	\end{eqnarray}
	such that the energy equation \eqref{eq:energy-eq} satisfies.
	% \begin{eqnarray}
		% \label{eq:phi_and_phidot}
		% \frac{\dot{\phi}^2}{6 }\equiv H^2 \sin^2{\theta}, \quad \text{and}\quad \frac{V}{3}\equiv H^2\cos^2{\theta}.  
		% \end{eqnarray}
	Differentiating with respect to the cosmic time $t$ and after re-arranging terms, we get,
	\begin{eqnarray}
		\label{eq:gen_thetadot}
		&&\dot{\theta} = \frac{V_{,\phi}}{\sqrt{2 V}} - \frac{3}{2}H \sin{2\theta},\\
		\label{eq: first_slow_roll_in_theta}
		&&\dot{H} = - 3 H^2 \sin^2{\theta}.
	\end{eqnarray}
	Such choice of orientation simply implies that, as $\theta \ll 1$, $\dot{\phi}$ is negative and significantly smaller than the Hubble parameter, indicating the slow-roll regime. On the other hand, $\theta=\theta_{\rm{end}} = \sin^{-1}\left(\frac{1}{\sqrt{3}}\right)$ defines the exact epoch of end of inflation, for $\theta = (2n + 1)\frac{\pi}{2}, n \in N,$ the potential vanishes, and this corresponds to the bottom at the potential, and for $\theta = n \pi,~n \in N,$ the field velocity is zero, and the field reaches the peak of the potential. Unlike the slow-roll approximation, these values are exact, which is one of the key reasons for using such formalism.
	%which subsequently gives us
	%\begin{eqnarray}
	%\label{eq:gen_H}
	%H= \frac{1}{3\int{\frac{\sin^2{\theta}}{\dot{\theta}}} + C_H}, \quad C_H \equiv %\text{Constant.}
	%\end{eqnarray}
	
	Therefore, instead of using the background equations \eqref{eq:energy-eq}, \eqref{eq:acc-eq} and \eqref{eq:phi}, here we analyze the full solution of the system by solving the Eqs. \eqref{eq:gen_thetadot} and \eqref{eq: first_slow_roll_in_theta}. Keeping that in mind, let us again consider the case of chaotic inflation. Then,  Eq. \eqref{eq:gen_thetadot} can be rewritten as
	\begin{eqnarray}
		\label{eq:theta_dot_chaotic_gen}
		\dot{\theta}= m-\frac{3}{2}H \sin{2\theta}.
	\end{eqnarray}
As mentioned earlier, after the end of inflation, the Hubble parameter falls significantly below effective mass, i.e., $H \ll m.$ As a consequence, during reheating, the Eq. \eqref{eq:gen_thetadot} can be approximated and solved as
	
	\begin{eqnarray}
		\label{eq:reheating_thetadot_chaotic}
		&&\dot{\theta}  \simeq m,\qquad\qquad \theta = \theta_{\rm{0}} + m (t-t_{\rm{0}}) ,
	\end{eqnarray}
	where $\theta(t=t_{\rm{0}})=\theta_{\rm{0}}$. Integrating Eq. \eqref{eq: first_slow_roll_in_theta}, we can write the solution of Hubble parameter $H$ as a function of $\theta$ as
	\begin{eqnarray}
		H = \frac{ H_{\rm{0}} }{1 + \frac{3 H_{\rm{0}}}{4 m}\left( 2 (\theta - \theta_{\rm{0}})-   (\sin 2 \theta - \sin 2\theta_{\rm{0}})\right)},
	\end{eqnarray}
	% \begin{eqnarray}
		% % -\frac{H_{,\theta}}{H^2}=\frac{3 \sin^2{\theta}}{\dot{\theta}},\quad
		% H=\frac{m }{3(\frac{\theta}{2}-\frac{1}{4}\sin{2\theta})+m ~C},
		% \end{eqnarray}
	% where $C$ is a constant of integration. During reheating $\theta\gg1$, the above equation can be rewritten as
	where $ H = H_{\rm{0}}$ at $\theta = \theta_{\rm{0}}.$ In the deep oscillating stage, $\theta \gg \theta_{\rm{0}}$ and $t \gg t_{\rm{0}}$, which  can approximate the above solution to
	\begin{eqnarray}\label{eq:H-rehet-theta-limit}
		H \simeq  \frac{2 m}{3 \theta} \left(1 + \frac{\sin 2 \theta}{2 \theta}\right),
	\end{eqnarray}
	where $\theta_{\rm{0}}$ is chosen at any time at the bottom of the potential. Substituting the solution of $\theta $ from Eq. \eqref{eq:reheating_thetadot_chaotic} in above equation we get,
	\begin{eqnarray}
		% -\frac{H_{,\theta}}{H^2}=\frac{3 \sin^2{\theta}}{\dot{\theta}},\quad
		H \simeq \frac{2 }{3 t} \left(1 + \frac{\sin 2 m t}{2 m t}\right).
	\end{eqnarray}
	Notice that the time average of the Hubble parameter behaves as $2/(3t)$, i.e., like a dust-matter-dominated solution with the effective equation of state $w_{\rm re} = 0$. The corresponding solution for the field $\phi$ in this regime can be written as
	\begin{eqnarray}
		\phi\simeq\frac{2 \sqrt{2}}{\sqrt{3}m t}\cos m t\\
		\dot{\phi}\simeq-\frac{2 \sqrt{2}}{\sqrt{3}t}\sin m t
	\end{eqnarray}
	Also, the first and the second slow-roll parameters can be written as
	\begin{eqnarray}
		\epsilon_{\rm{1}} &=& 3 \sin^2{ m t},\\
		\epsilon_{\rm{2}} &=& 3 m t \cot{ m t}.
	\end{eqnarray}
	This is the complete solution for the reheating era for the case of the chaotic inflationary model.
	
	Please note that the inflation ends at $\theta_{\rm{end}}$, and the field, for the first time, reaches the bottom of the potential, making the first oscillation at $\theta = \pi/2.$ At and around this stage, $H \sim m.$ Therefore, the reheating solution \eqref{eq:reheating_thetadot_chaotic} and the solutions thereafter cannot be trusted as the above solutions are obtained using the approximation $H \ll m$. Only after a few oscillations, when $H$ falls significantly below the mass of the potential $m,$ the reheating solutions asymptotically merge with the solutions given above. 
	
	To summarize, in this section, for the chaotic inflation model, we derive the dynamics of the Universe in two distinct regimes. For $|\phi| \gg 1$, slow-roll conditions are met, and using these conditions, we derive the slow-roll dynamics, which leads to a quasi-exponential scale factor solution. For $|\phi| \ll 1$, however, the field oscillates around the minimum of the potential and decays into other particles, referred to as the reheating epoch, and using $H \ll m$ approximations, we also obtain the asymptotic solution in this epoch. The two approaches to achieving these two extreme solutions are also entirely distinct. As previously stated, the solution when $H \sim m$ is still not well understood, and the two solutions given above do not justify around this regime. And because the methodologies are distinct, extrapolating these two solutions into a single solution is also exceedingly challenging. In the following section, we will demonstrate that this is, in fact, possible if we contemplate a single method of solving these two regimes, which in our case is identical to the method used to solve the reheating era, which is characterized by the variable $\{\theta, H\}$.

	\section{extending the phase space solution method in slow-roll regime for chaotic inflation}\label{sec:chaotic-slow-roll}
	
	%%%%%%%%%%%%%%%%%%%%%%%%%%%%%%%%%%%%%%%%%%%
	%%%%%%%%%%%%%%%%%%%%%%%%%%%%%%%%%%%%%%%%%%%
	%%%%%%%%%%%%%%%%%%%%%%%%%%%%%%%%%%%%%%%%%%%
	
	%In the previous section, we discussed the reheating solution using the $\theta$ coordinate which defines the phase space rotation. Previously, we also obtained the slow-roll solutions using the slow-roll parameters. In this section, we will show how to obtain slow-roll solutions using the method used in the reheating era. Then, we will also show how to extend the solution for the entire inflation-reheating era for the chaotic inflation model.
	
	%\subsection{Slow-roll}
	Let us now focus on the method to analyze the evolution of the Universe during the slow roll. As mentioned earlier, during this epoch, $\theta$ is small, and as a consequence, using Eq. \eqref{eq: first_slow_roll_in_theta}, we can approximate the first slow-roll parameter as
	\begin{eqnarray}
		\label{eq:first-slow-roll-theta-inflation}
		\epsilon_{\rm{1}} \equiv -\frac{\dot{H}}{H^2} \simeq 3\theta^2.
	\end{eqnarray}
	Using Eq. \eqref{eq:slow_roll_in_V}, one can immediately obtain the relation between the variable $\theta$ and the scalar field $\phi$ as 
	% \begin{eqnarray}
		%  \frac{V_{,\phi}}{\sqrt{2 V}}\simeq 3 H \theta
		% \end{eqnarray}
	\begin{eqnarray}
		\label{eq:slow_roll_theta}
		\theta \simeq \frac{V_{,\phi}}{\sqrt{6}V},
	\end{eqnarray}
	which, in turn, leads to
		
	\begin{eqnarray}
		\label{eq:slow_roll_thetadot}
		% \dot{\theta}&\simeq& \frac{1}{\sqrt{6}}d_\phi\Big\{ \frac{V_{,\phi}}{V}\Big\} \dot{\phi}=-\frac{1}{3\sqrt{2}}\Big\{\frac{V_{,\phi \phi}}{V}-\frac{V_{,\phi}^2}{V^2}\Big\}\frac{V_{,\phi}}{\sqrt{V}}
		\dot{\theta} \simeq -\frac{1}{3\sqrt{2}}\left(\frac{V_{,\phi \phi}}{V}-\frac{V_{,\phi}^2}{V^2}\right)\frac{V_{,\phi}}{\sqrt{V}}.
	\end{eqnarray}
	In the case of chaotic inflation, the above equations take the following form:
	
	\begin{eqnarray}\label{eq:slow_roll_theta_chaotic}
		\theta\simeq\sqrt{\frac{2}{3}}\frac{1}{\phi},\qquad
		\dot{\theta}\simeq m \theta^2.
		% \phi=\sqrt{\frac{2}{3}}\frac{1}{\theta}
	\end{eqnarray}
	% Eq.\eqref{eq:slow_roll_thetadot} takes the form,
	% \begin{eqnarray}
		% \dot{\theta}&\simeq& m \theta^2\\
		% \theta&\simeq& \frac{1}{C-m t}
		% \end{eqnarray}
	Note that, $\phi \gg 1$ leads to $\theta \ll 1,$ and also $\dot{\theta} \ll 1,$ which make the above assumptions self-consistent. We now can integrate the above equation as:
	
	\begin{eqnarray}\label{eq:theta_slowroll_chaotic}
		% \dot{\theta}&\simeq& m \theta^2\\
		\theta&\simeq& \frac{\sqrt{2}}{\sqrt{3}\phi_{i}-\sqrt{2}m t},
	\end{eqnarray}
	where $\phi(t = 0) \equiv  \phi_{i}.$ We can also integrate Eq. \eqref{eq:first-slow-roll-theta-inflation} and obtain the relation between $H$ and $\theta$ as
	\begin{eqnarray}\label{eq:H_slowroll_chaotic}
		% -\frac{H_{,\theta}}{H^2}=\frac{3 \sin^2{\theta}}{\dot{\theta}},\\
		H \simeq \frac{m}{3\theta},
		% =\frac{m\phi}{\sqrt{6}-m\phi C}.
	\end{eqnarray}
	where we use the initial condition for chaotic inflation $H(\theta \rightarrow 0) \rightarrow \infty.$ It can now be seen that Eqs. \eqref{eq:theta_slowroll_chaotic} and \eqref{eq:H_slowroll_chaotic} are in agreement with the Eqs. \eqref{eq:slow_roll_chaotic} and \eqref{eq:slow_roll_chaotic_H}.
	
	Please note that since we now solve the system using the variables $\{\theta, H\}$ even during slow-roll regime, the fundamental difference in the dynamics of the chaotic inflationary model comes only in the expression $\dot{\theta}$: i.e., during reheating, it was simply $m,$ a constant, whereas, during slow-roll, it takes the form $m \theta^2.$ In the following section, we will propose a method for obtaining the entire solution using this information as an advantage.
	\section{Proposed full solution for chaotic inflation}\label{sec:full-chaotic}
	For chaotic inflation, we have discussed the dynamics of the Universe in two different regimes, i.e., during the slow-roll and reheating era. Let us just summarize the method in brief. Instead of Eqs. \eqref{eq:energy-eq}, \eqref{eq:acc-eq}, and \eqref{eq:phi}, expressed in $\phi$ and $\dot{\phi},$ we redefine these equations in terms of the variable $\theta$ and $H,$ and equivalently, obtain two generalized equations \eqref{eq:gen_thetadot} and \eqref{eq: first_slow_roll_in_theta}. Then, in the case of either inflation or reheating, we express $\dot{\theta}$ as a function of $\theta$, i.e., during reheating epoch, $\dot{\theta}$ is constant, whereas, during the slow-roll phase, it goes as $\propto \theta^2$ (\viz Eqs. \eqref{eq:reheating_thetadot_chaotic} and \eqref{eq:slow_roll_theta_chaotic}):

	\begin{eqnarray}
		\dot{\theta}\simeq\left\{\begin{array}{cc}{m \theta^2} & {~~~ \text{Slow-roll}} \\ {m} & {~~~~ \text{Reheating}} \\ \end{array}\right.
	\end{eqnarray}
	
	By solving these equations together with Equation \eqref{eq: first_slow_roll_in_theta}, we obtain the dynamics in these two distinct regimes. Please note, however, that, as previously remarked, these approximations do not hold at the end of the inflation era and the beginning of the reheating era, and the solution can only be completed if we know how $\dot{\theta}$ behaves in this adjacent era.
	
	Therefore, to obtain a complete solution from slow-roll inflation to reheating, we require a solution in which $\dot{\theta}$ behaves as $m \theta^2$ for $\theta \ll 1$ and as $m$ for $\theta \gg 1$, with a seamless transition between these two solutions. Without worrying about the actual solutions, one can make an intellectual conjecture as to the form of such functions, and the possibilities are limitless. In this paper, we find one basic yet effective form such that the solution can be solved analytically by simple functions with the form:

	\begin{eqnarray}\label{eq:thetadot_chaotic_full}
		\dot{\theta}&=& \frac{m \theta^2}{1+ \theta^2}.
	\end{eqnarray}

	Note that inflation ends exactly at $\theta_{\rm{end}} \simeq 0.6$, and the field, for the first time, reaches the bottom of the potential at $\theta = \pi/2 \sim 1.57$. Therefore, one can verify that, for the inflationary as well as the reheating solution, the above assumption depicts near-accurate dynamics of $\dot{\theta}$ with the solution of $\theta$ as
	
	\begin{eqnarray}\label{eq:sol-theta-full-chaotic}
		\theta = \left(\theta_{i}^2 + m \theta_{i}t-1\right) + \sqrt{4\theta_{i}^2+\left(\theta_{i}^2 + m \theta_{i}t-1\right)^2}
		% &&\epsilon_{\rm{1}}= 3 \sin^2\left( \frac{1}{2\theta_{i}}\left(\sqrt{4\theta_{i}^2+\zeta^2}+\zeta\right)\right),\\
		% \quad\zeta=\left(m \theta_{i}(t-t_{i})-1+\theta_{i}^2\right)
	\end{eqnarray}
	where, $\theta(t = 0)\equiv \theta_{i}\equiv\sqrt{\frac{2}{3\phi_{i}}}$, is the initial condition, chosen during the deep slow-roll regime.
	
	\begin{figure}[t!]
		\includegraphics[width=0.495\textwidth]{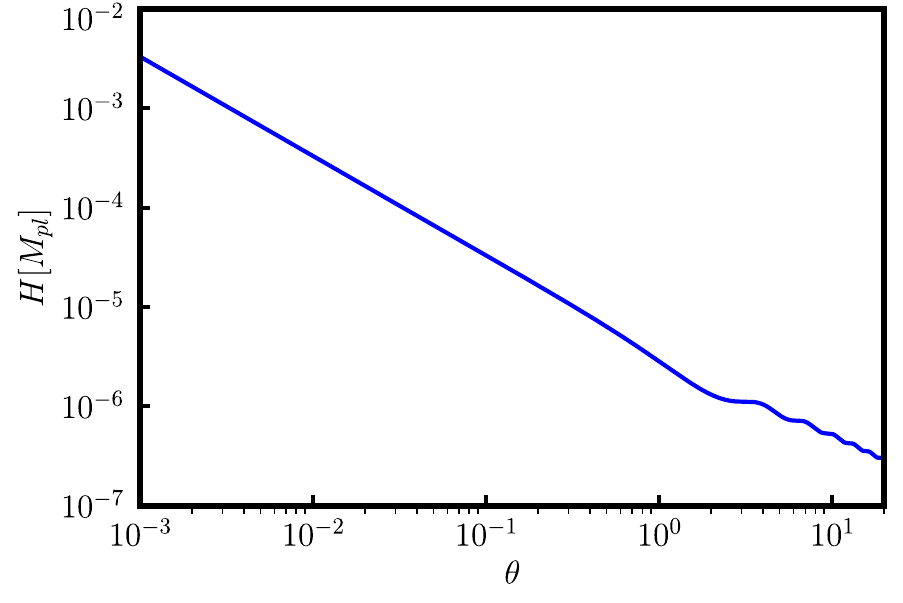}
		\includegraphics[width=0.495\textwidth]{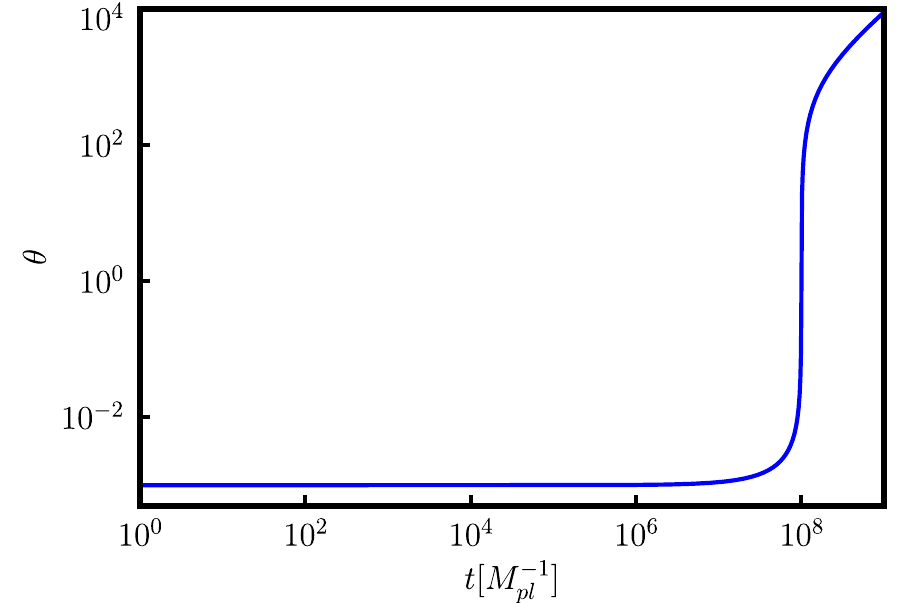}
		\caption{Chaotic inflaion: We plot the Hubble parameter $H$ as a function of $\theta$ (top), and $\theta$ as a function of cosmic time $t$ (bottom).
			\label{fig:chaotic_H_vs_theta}
		}\vspace{-4mm}
	\end{figure}
	Now that we know the solution of $\dot{\theta},$ one can again use the equation related to the first slow-roll parameter, i.e., Eq. \eqref{eq: first_slow_roll_in_theta} and solve the Hubble parameter and the subsequent dynamics. Similar to the previous scenario, integrating Eq. \eqref{eq: first_slow_roll_in_theta} and using Eq. \eqref{eq:thetadot_chaotic_full}, evolution of the $H$ as function of $\theta$ can be obtained for the above model as

	\begin{eqnarray}\label{eq:sol-H-theta-chaotic}
		H = \frac{4m\theta }{3(2 (\theta ^2-1)- \theta  \sin  2 \theta + 2 \cos 2 \theta + 4 \theta\  \text{Si}(2 \theta ))},
	\end{eqnarray}
	where, once again, we use the initial condition, $H(\theta \rightarrow0) \rightarrow \infty $ for the chaotic inflation, and Si$(x)$ is Sine Integral function. Note that $\text{Si}(x)\rightarrow x$ for $x\rightarrow 0$, and as a result, under the limit $\theta \ll 1,$ one can verify that the above solution coincides with the solution of slow-roll chaotic inflationary solution given in Eq. \eqref{eq:slow_roll_chaotic_H}. On the other extreme limit,  $\text{Si}(x)\rightarrow \pi/2$ for $x \gg 1,$ and thus, for $\theta \gg 1,$ the above solution coincides with the reheating solution \eqref{eq:H-rehet-theta-limit}. Therefore, the above solution of the Hubble parameter is consistent with both the slow-roll as well as the reheating solutions discussed in the previous section. Similarly, the general solutions of $\phi$ and $\dot{\phi}$, by using Eq \eqref{eq:phi_and_phidot}, can be written as:
	\begin{eqnarray}
		\phi &=& \frac{4\sqrt{6}\theta \cos \theta }{\left(6 (\theta ^2-1)-3 \theta  \sin  2 \theta + 6 \cos 2 \theta + 12 \theta\  \text{Si}(2 \theta )\right)},\nonumber \\
		~\\
		\dot{\phi} &=& -\frac{4 \sqrt{6} m \theta \sin \theta }{\left(6 (\theta ^2-1)-3 \theta  \sin  2 \theta + 6 \cos 2 \theta + 12 \theta\  \text{Si}(2 \theta )\right)},\nonumber\\
	\end{eqnarray}
	and the two slow-roll parameters can be expressed as:
	\begin{eqnarray}
	\epsilon_{\rm{1}} = 3 \sin^2 \theta,\qquad
	\epsilon_{\rm{2}} = \frac{ 3\theta \cot \theta (2 (\theta ^2-1+\cos 2\theta)+ \theta(4\text{Si}(2 \theta )- \sin  2 \theta))}{2 (1 + \theta^2)}.
	\end{eqnarray}

	\begin{figure}[t]
			\includegraphics[width=0.495\textwidth]{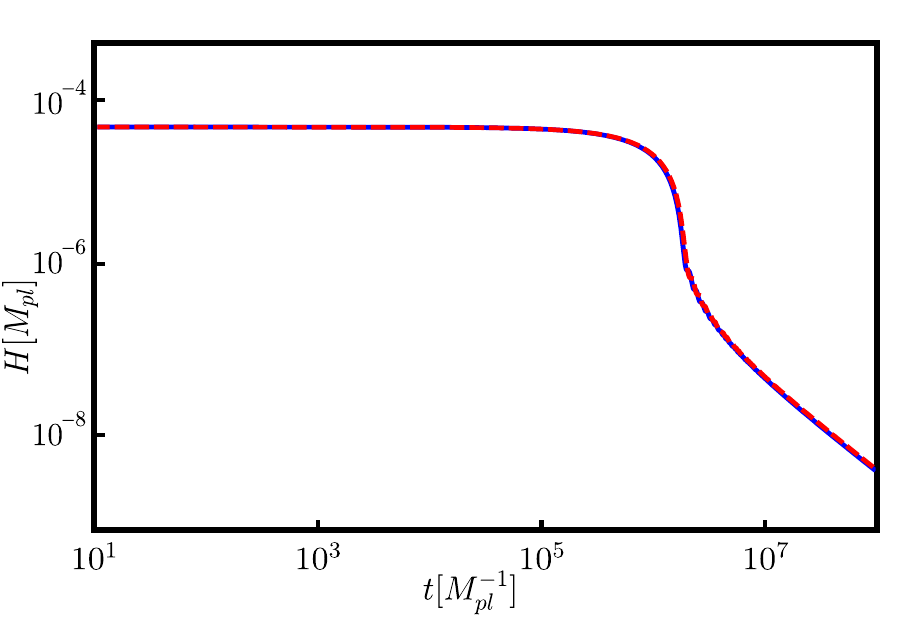}
			\includegraphics[width=0.495\textwidth]{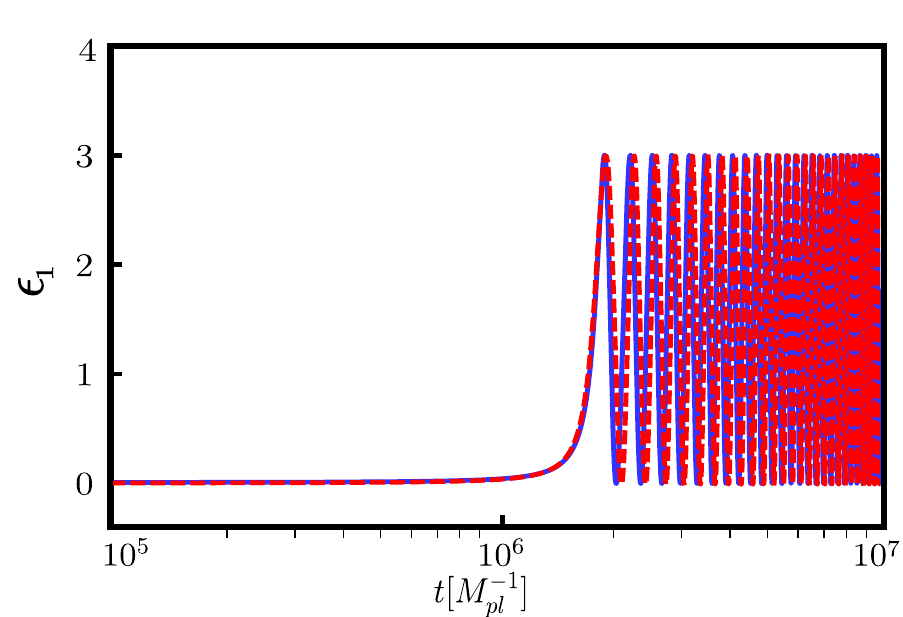}
			\includegraphics[width=0.495\textwidth]{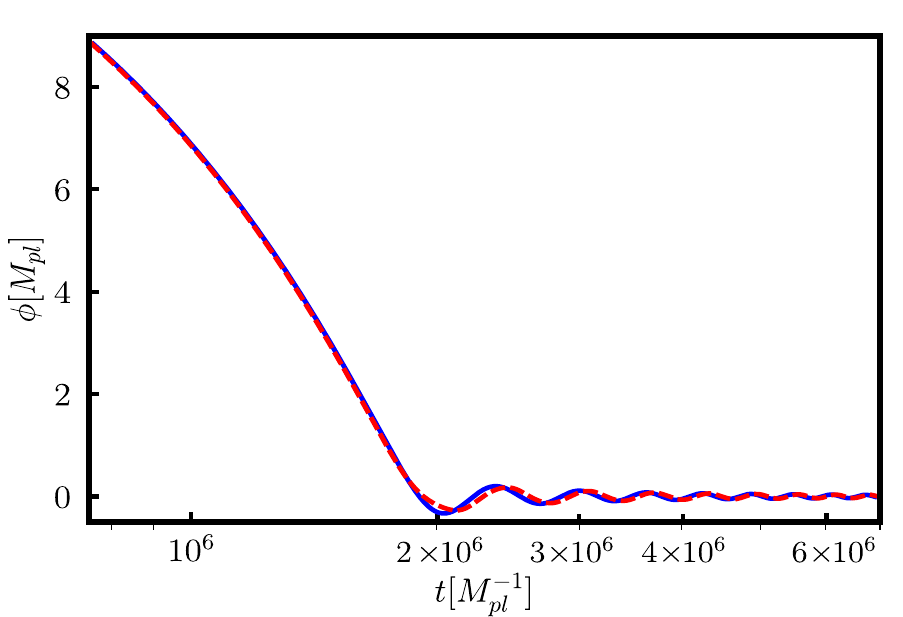}
			\includegraphics[width=0.495\textwidth]{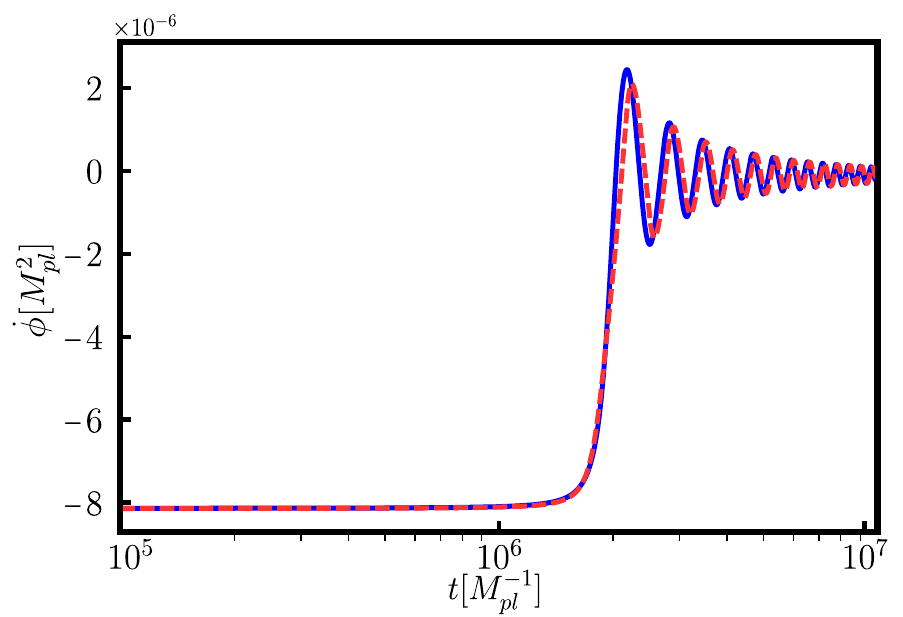}
		\caption{Chaotic inflation: We plot the Hubble parameter $H$ (top-left), first slow-roll parameter $\epsilon_{\rm{1}}$ (top-right), $\phi$ (bottom-left), and $\dot{\phi}$ (bottom-right) as functions of cosmic time $t$ both numerically (dotted red) and analytically (solid blue) and show that the analytical solution provides a good level of accuracy of evaluating the background dynamics.
			\label{fig:chaotic_H_eps_phi_dphi_vs_t}
			\vspace{-3mm}
		}
	\end{figure}

	Please note that the effect of the Si$(x)$ function may appear to be irrelevant. Nonetheless, this function plays a crucial role in the transition between the two epochs and provides greater precision; therefore, it cannot be neglected.
	Let us now discuss the impact of the full solution, which is demonstrated in Fig. \ref{fig:chaotic_H_vs_theta}. 
	As can be seen, since the expressions of all variables are now given in terms of theta, we know the precise values of these variables in each physical instance, as specified by the variable $\theta$. Consider, for example, the variable Hubble parameter $H$ given in Eq. \eqref{eq:sol-H-theta-chaotic}. At $\theta = \sin^{-1}(1/\sqrt{3})$, inflation ends precisely, and the above analytical solution yields $H_{\rm{end}} \simeq 0.503 m$ at this level. In contrast, using the slow-roll approximation, we previously obtained $H_{\rm{end}} \simeq m/\sqrt{3} \simeq 0.577 m$ using analytic techniques. Using numerical simulations, we determine that $H_{\rm{end}} \simeq 0.504m,$ which demonstrates that our method provides a much higher degree of accuracy. On the other hand, $\theta = \pi/2$ denotes when the field reaches the bottom of the potential for the first time. Using our approach, it is now obvious that $H \simeq 0.167 m.$ For the second time, it reaches the bottom, $H \simeq 0.087 m,$ for third time, it is $H \simeq 0.061 m,$ and so on.  On the other hand, when the field reaches its first maxima, $H \simeq 0.112 m,$ at the second maxima, $H \simeq 0.072 m,$ for the third, it is $0.053 m,$ and so on. 
	\begin{figure}
		\centering
		\includegraphics[width=0.8\textwidth]{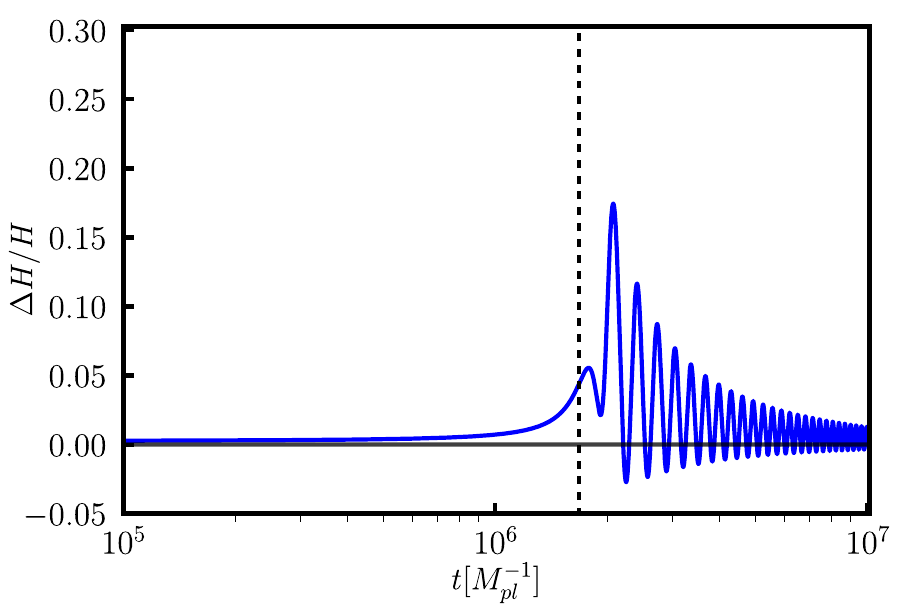}
		\caption{Chaotic inflation: we plot the relative difference between the numerical and analytical (proposed solution) of Hubble parameter $H$ as a function of cosmic time $t$ and show that the analytical solution provides a good level of accuracy of evaluating the background dynamics. Here black (dashed) line corresponds to epoch of end of inflation.
		\label{fig: relative_diff}}
		\vspace{-5mm}
	\end{figure}
	
	 In fact, using the above expressions \eqref{eq:sol-theta-full-chaotic} and \eqref{eq:sol-H-theta-chaotic}, once we know the initial conditions (i.e., $\phi$ and $\dot{\phi}$), we can easily evaluate $\theta,$ and subsequently, the value of the Hubble parameter. Other variables such as $\epsilon_{\rm{1}}$, $\phi$, and $\dot{\phi}$  are then straightforward to evaluate. Thus, our analysis, without the need for numerical simulations, analytically and with great accuracy, reflects how the Universe evolves with time during any epoch, be it inflation or the reheating epoch (see Figs. \ref{fig:chaotic_H_eps_phi_dphi_vs_t}, \ref{fig: relative_diff} as well as Table \ref{table:1}). 
	 
	 To grasp the degree of accuracy, let us look at the Fig. \ref{fig: relative_diff}, which illustrates the relative difference between the numerical and our proposed analytical solutions to the Hubble Parameter. Deep during the slow-roll phase, this difference is negligible with the relative error is being less than $1 \%$ ($\sim 0.2 \%$ at the pivot scale), whereas the only prominent difference observed around the end of inflation of $\sim 15 \%$, which eventually dimnishes/reduces during the oscillatary phase. It implies that our proposed analytical method depicts a very good approximation of the entire slow-roll as well as reheating dynamics relative to the slow-roll and oscillatory reheating approximations. This is one of the main results of this work.

		\begin{center}
			\begin{table}[t]
				\begin{tabular}{ |c|c|c|c|c|c|c|} 
					\hline&&&&&&\\
					 Epoch & 
					$\hspace{0.5cm}\theta\hspace{0.5cm}$&    $\hspace{0.5cm}\phi\hspace{0.5cm}$ & $\hspace{0.5cm}\dot{\phi}\hspace{0.5cm}$  &
					$\hspace{0.5cm}H\hspace{0.5cm}$  &  $\hspace{0.5cm}\epsilon_{\rm{1}}\hspace{0.5cm}$ &  $\hspace{0.5cm}\epsilon_{\rm{2}}\hspace{0.5cm}$   \\[0.8ex] 
					\hline\hline
					Scale factor $a(t) \rightarrow 0$ & $0$ &  $\infty$ & $0.816 \ m $ &$\infty$ & 0 & 0 \\
					\hline
					End of inflation&$\sin^{-1}(1/\sqrt{3})$ &  $\ 1.006$ & $-0.711\ m$ &$0.503\ m$ & $1 $ & $1.544$ \\
					\hline
					First minimum &$\pi/2$ &  $0$ & $-0.408\ m$ &$0.167\ m$ & $3 $ & $0$ \\
					\hline
					First maximum &$\pi$ &  $-0.273$ & $0$ & $0.111\ m$ &$0$ & $\infty$\\
					\hline
					Second minimum &$3\pi/2$ & $0$ & $0.213\ m$ & $0.087\ m$ & $3 $ & $0$ \\
					\hline
					Second maximum &$\pi$ &  $0.176$ & $0$ &$0.072\ m$ & $0$ & $\infty$\\
					\hline
				\end{tabular}
				\caption{Chaotic inflation: Different values of $\phi$, $\dot{\phi}$, $H,~\epsilon_{\rm{1}}$ and $\epsilon_{\rm{2}}$ for different values of $\theta$.}
				\label{table:1}
			\end{table}
		\end{center}

\section{Extended general solution}\label{sec:gen sol}

In this section, we will extend the solution for chaotic inflation to other inflationary models. As we saw earlier, during the slow-roll stage, in the case of chaotic inflation, $\dot{\theta} \propto \theta^2$, in the generalized scenario, we can extend the exponent factor from two to any arbitrary real positive number $n$. During reheating, however, we already know that if the potential behaves nearly as $V(\phi) \propto \phi^2$, then $\dot{\theta}$ remains constant during this time period. For simplicity and to derive simple analytical expressions, we also want to maintain this relationship in the generalized scenario, i.e., near the bottom of the potential, we want it to behave similarly to $\propto \phi^2$. A prime example, which will be discussed in the following section, is the Starobinsky inflation, where the potential is nearly flat during the slow-roll regime but behaves as $\phi^2$ near the bottom of the potential. Hence, for generalized inflationary models, $\dot{\theta}$ can be expressed as

\begin{eqnarray}
	\dot{\theta}\simeq\left\{\begin{array}{cc}{\mu\theta^n} & {~~~ \text{Slow-roll}} \\ {\nu} & {~~~~ \text{Reheating}} \\ \end{array}\right.
\end{eqnarray}
where $\mu, ~\nu$, and $n$ are all positive constants that can be correlated with the model parameters, i.e., the inflationary potential. Here, we assume the potential's nature is simple and that the transition from the slow roll to the reheating scenario is seamless. Any feature of the potential or deviation from the slow-roll, such as ultra slow-roll, is not taken into account, as $\dot{\theta}$ may differ from the above expression in such cases.

Similar to chaotic inflation discussed in the previous section, we now can combine both cases and propose the general solution of $\dot{\theta}$ as a function of $\theta$ as 
\begin{eqnarray}
	\label{eq:gen_thetadot-as-theta}
	\dot{\theta} = \frac{\mu \theta^n}{1+ \frac{\mu}{\nu} \theta^{n}}.
\end{eqnarray}
The above equation can be integrated to get the solution of $\theta$ in terms of the cosmic time as
\begin{eqnarray}\label{eq:gen-theta-sol-t}
	\mu (1-n) \theta +\nu \theta^{1-n} = C_{\rm{1}} + \mu \nu(1-n) t,
\end{eqnarray}
where $C_{\rm{1}} \equiv (1-n) \mu \theta_{i}+\nu \theta_{i}^{1-n}$ is the constant of integration, and $\theta(t = 0) \equiv \theta_{i}$. The dependence of $\theta_{i}$ on $\phi_{i}$ depends on the form of potential. Again, using Eq. \eqref{eq: first_slow_roll_in_theta} along with the above Eq. \eqref{eq:gen_thetadot-as-theta} and other equations, we now can obtain the solution corresponding to $H$, $\epsilon_{\rm{1}}$, $\epsilon_{\rm{2}}$, $\phi$, and $\dot{\phi}$ as a function of $\theta$ as

	\begin{eqnarray}\label{eq:gen-H-theta}
		&&H=\frac{4 \mu  \nu  (n-1) \theta ^n}{\mu  (n-1)  (4 \nu C_{\rm{2}} +6 \theta -\ 3 \sin 2 \theta )\ \theta ^n + 3  \nu  (n-1) \left(E_{n}(2 i \theta ) + E_{n}(-2 i
			\theta )\right)\theta - 6\nu \theta},\nonumber\\&&\\
		\label{eq:gen-eps1-theta}
	&&	\epsilon_{\rm{1}}=3 \sin^2\theta,\\
		\label{eq:gen-eps2-theta}
		&&\epsilon_{\rm{2}}= \frac{\mu  (n-1)  (4 \nu C_{\rm{2}} +6 \theta -\ 3 \sin 2 \theta )\ \theta ^n + 3  \nu  (n-1) \left(E_{n}(2 i \theta ) + E_{n}(-2 i
			\theta )\right)\theta - 6\nu \theta}{4 (n-1) (\nu+ \mu \theta^{n})}\cot \theta,\nonumber\\&&\\
		\label{eq:gen-phidot-theta}
		&&\dot{\phi}=-\frac{4\sqrt{6} \mu  \nu  (n-1) \theta ^n \sin \theta}{\mu  (n-1)  (4 \nu C_{\rm{2}} +6 \theta -\ 3 \sin 2 \theta )\ \theta ^n + 3  \nu  (n-1) \left(E_{n}(2 i \theta ) + E_{n}(-2 i
			\theta )\right)\theta - 6\nu \theta},\nonumber\\&&\\
		\label{eq:gen-pot-theta}
		&&V(\phi)=\frac{48 \mu^2  \nu^2  (n-1)^2 \theta ^2n \cos^2\theta}{\left(\mu  (n-1)  (4 \nu C_{\rm{2}} +6 \theta -\ 3 \sin 2 \theta )\ \theta ^n + 3  \nu  (n-1) \left(E_{n}(2 i \theta ) + E_{n}(-2 i
			\theta )\right)\theta - 6\nu \theta\right)^2},\nonumber\\&&
	\end{eqnarray}

\noindent where $E_{n}(x)$ is the exponential integral function, and $C_{\rm{2}}$ is the integration constant, which again depends on the shape of the potential.

The background variable solutions mentioned above describe the evolution not only during the slow-roll and reheating phases but also during the transition phase. Additionally, it is model-independent, meaning that it may be used with both small and large field models, and the values of $\mu~, \nu$, and $n$ are determined by the model parameters associated with the potential. As a result, it represents the comprehensive, model-independent solution of all the dynamical variables during the entire evolution from slow-roll to reheating, which is the main outcome of this article.

Before proceeding into the next section, let us now discuss the parameters $C_{\rm{2}}$ and $n.$ In order to determine the value or the range of it, let us again consider the slow-roll regime. During slow-roll, i.e., for $\theta \ll 1,$ the Hubble and the slow-roll parameters take the form:

\begin{eqnarray}\label{eq:gen-Hubble-slow-roll-theta}
	&&H = \frac{1}{C_{\rm{2}} + \frac{3}{(3 - n)\mu}\ \theta^{3 - n}}.\\
	\label{eq:gen-eps1-slow-roll-theta}
	&&\epsilon_{\rm{1}}=3 \theta^2
	\label{eq:gen-eps2-slow-roll-theta}\\
	&&\epsilon_{\rm{2}}= 2 \mu \theta^{n-1}\left(C_{\rm{2}} + \frac{3}{(3 - n)\mu}\ \theta^{3 - n}\right)
\end{eqnarray}
Therefore, for $n < 3,$ the Hubble parameter approaches the value $1/C_{\rm{2}}$ when $\theta$ approaches zero. In such limit, in the case of large field models, $H \rightarrow \infty,$ whereas, for small field models, as $V(\phi)$ saturates to one value, say, $V_{\rm{0}}$ for $|\phi| \rightarrow \infty,$  $H$ takes the form $\sqrt{V_{\rm{0}}/3}.$  Therefore, $C_{\rm{2}}$ can be associated with these values for large and small field models as

\begin{eqnarray}\label{eq:C2}
	C_{\rm{2}} = \left\{\begin{aligned}
		& ~~~0, \quad ~~\ \text{large fields,}\\
		& \sqrt{\frac{3}{V_{\rm{0}}}}, \hspace{10pt} \text{small fields.}
	\end{aligned}\right.
\end{eqnarray}
The solution immediately translates to:

\begin{eqnarray}
	\dot{\phi} = -\frac{\sqrt{6} \theta}{C_{\rm{2}} + \frac{3}{(3 - n)\mu}\ \theta^{3 - n}},
\end{eqnarray}
and
\begin{eqnarray}
	V(\phi) \simeq \left\{\begin{aligned}
		&  \frac{\mu^2 (3 - n)^2}{3} \frac{1}{\theta^{6 - 2 n}}, \hspace{13pt} \text{large fields,}\\
		& \frac{3}{C_{\rm{2}}^2 + \frac{6 C_{\rm{2}}}{(3 - n)\mu}\theta^{3 - n}}, \quad \text{small fields.}
	\end{aligned}\right.
\end{eqnarray}
where $n < 3$ and $C_{\rm{2}}$ is given by Eq. \eqref{eq:C2}. The above expression, based on the functional form of the potential, leads to the functional dependence of $\phi$ over $\theta.$ One can also immediately notice that $n > 3$ is prohibited as $H$ becomes negative. At the same time, $n >0$ is required as $\dot{\theta} \rightarrow 0$ as $\theta \rightarrow 0.$ Therefore, the constraint on $n$ is

\begin{eqnarray}
	0 < n < 3.
\end{eqnarray}

Now that the generic solution has been provided, let's examine various inflationary models. There are typically two types of inflationary potential: large field potentials and small field potentials. Observations have, however already ruled out the possibility of large field inflationary potentials, such as the chaotic inflation. In contrast, among all small field models, we will discuss two types of models and determine their complete solutions in this paper.

\subsection{First kind of small field inflationary models}
\begin{figure}[h]
	\includegraphics[width=0.495\textwidth]{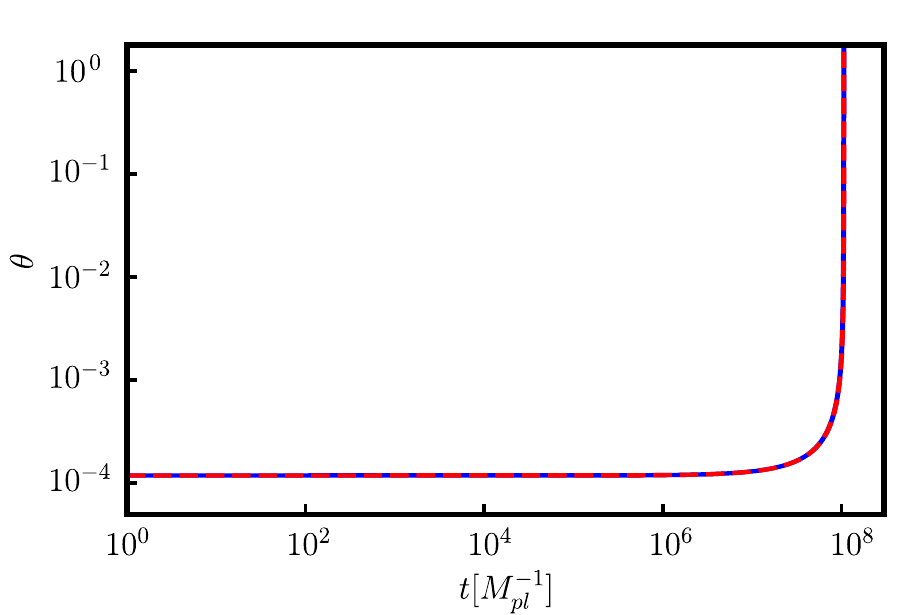}
	\includegraphics[width=0.495\textwidth]{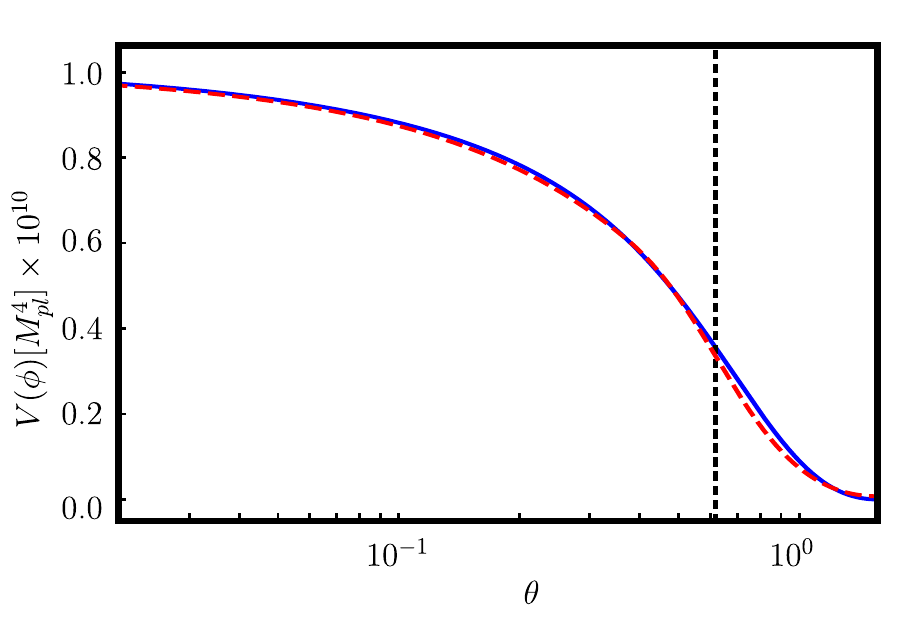}
	\caption{First kind of small field model: We plot of the evolution of $\theta$ as function of cosmic time $t$ (left) and potential $V(\phi)$ as function of $\theta$ (right) for the model parameters: $A=10^{-10}$, $m=10^{-5}$, and $\alpha=12$, both numerically (dotted-red) and analytically (solid-blue) for the domain $10^{-4} \leq \theta < \pi/2,$ i.e., from the deep inflationary to the bottom of the potential regime. These plots ensures that the analytical solution provides a good level of accuracy of evaluating the background dynamics.
		\label{fig:poly-n-alpha-plot}}
\end{figure}

\begin{figure}[h]
	\includegraphics[width=0.495\textwidth]{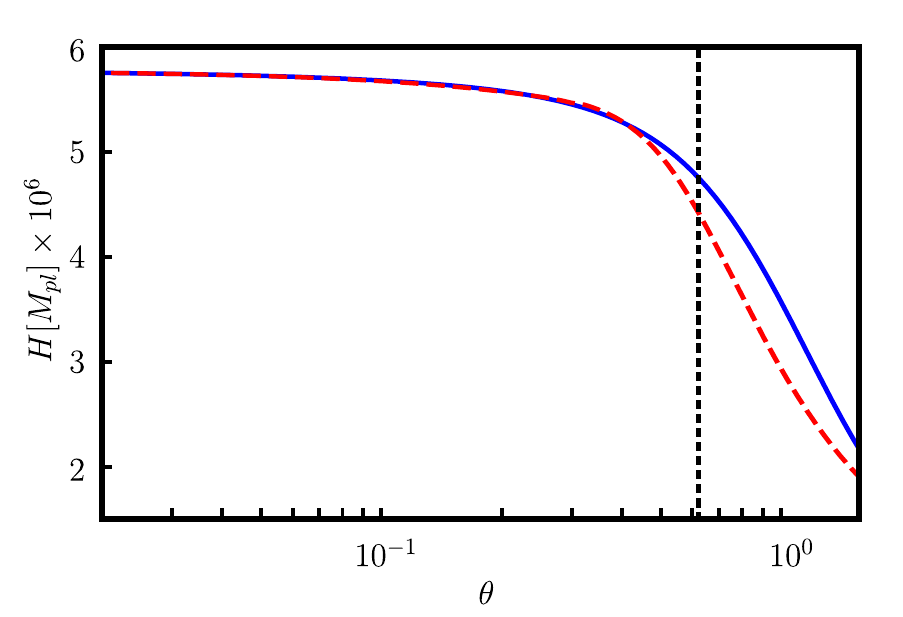}
	\includegraphics[width=0.495\textwidth]{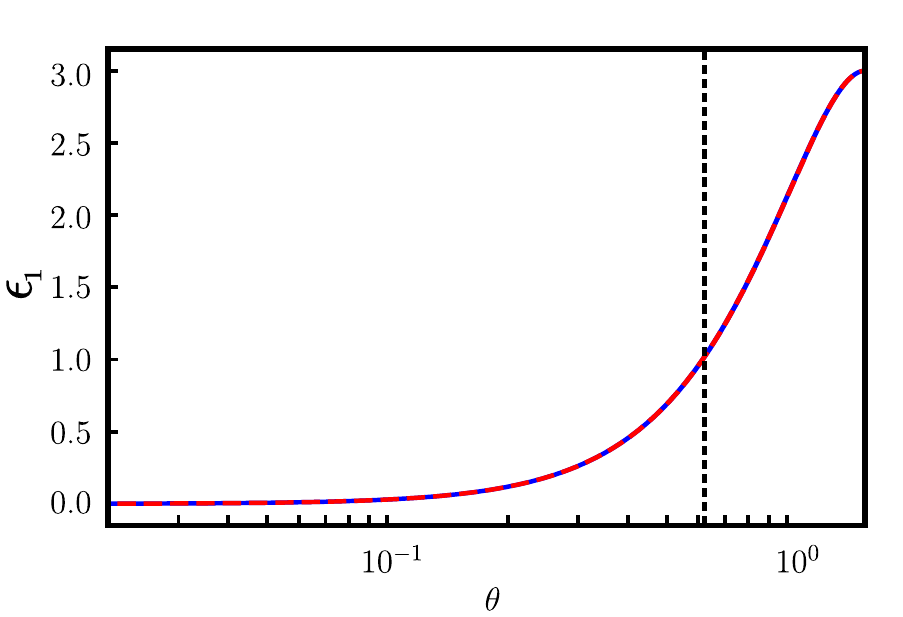}
	\includegraphics[width=0.495\textwidth]{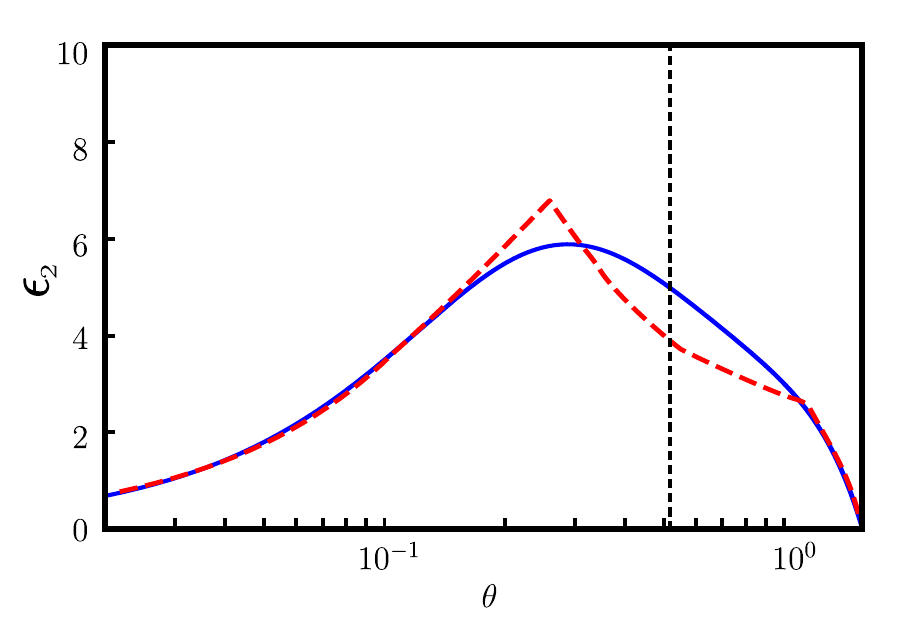}
	\includegraphics[width=0.495\textwidth]{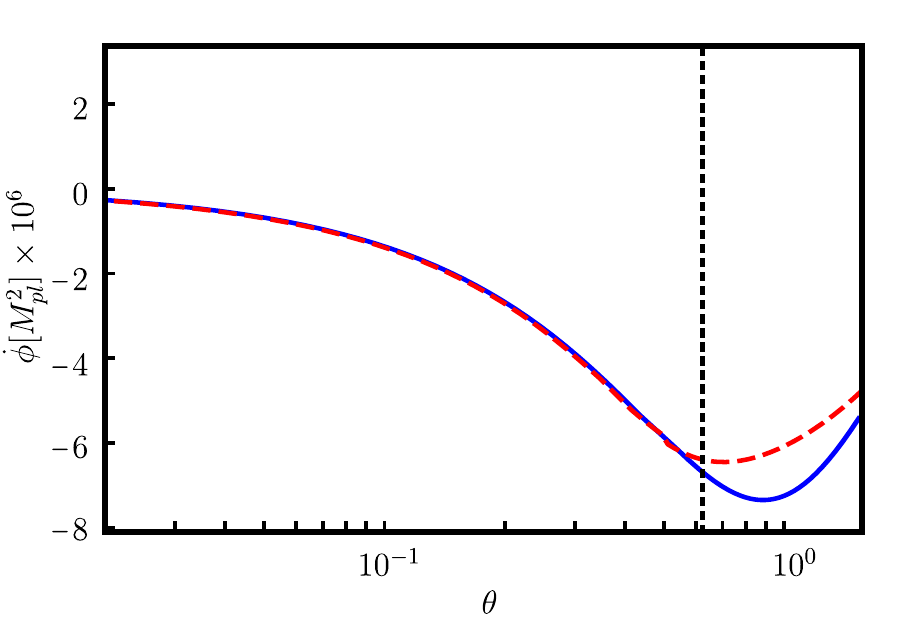}
	\caption{First kind of small field model: We plot the Hubble paramter $H$ (top-left), first slow-roll parameter $\epsilon_{\rm{1}}$ (top-right), second slow-roll parameter $\epsilon_{\rm{2}}$ (bottom-left), and $\dot{\phi}$ (bottom-right) as functions of cosmic time $t$ for the model parameters $A=10^{-10}$, $m=10^{-5}$, and $\alpha=12$, both numerically (dotted-red) and analytically (solid-blue) for the domain $10^{-4} \leq \theta < \pi/2,$ i.e., from the deep inflationary to the bottom of the potential regime.   These plots show that the analytical solution provides a good level of accuracy of evaluating the background dynamics.
		\label{fig:poly_H_eps_phi_dphi_vs_t}
	}
\end{figure}

This kind of model, during inflation, for $\phi \gg 1$ can be expressed as $V(\phi) \simeq A (1 - B \phi^{-\alpha}),~\alpha >0.$ Again, we assume the potential has a minimum at $\phi = 0,$ and around it, it has a form $V(\phi) \propto \phi^2$ for $\phi \ll 1.$ Therefore, the potential that we are interested in can be expressed as:

\begin{eqnarray}\label{eq:pot-poly}
	V(\phi)\simeq\left\{\begin{array}{cc}{A\left(1-B \phi^{-\alpha}\right)} & {~~~ \text{Slow-roll}} \\ {\frac{1}{2}m^2 \phi^2} & {~~~~ \text{Reheating}} \\ \end{array}\right.
\end{eqnarray}
where $A$, $B$, $m$, and $\alpha$  are constants. This kind of model is called the polynomial $\alpha$-attractor model \cite{Kallosh:2013yoa, Kallosh:2013tua, Galante:2014ifa, Carrasco:2015pla, Kallosh:2022feu, Bhattacharya:2022akq}.  
% {
	% \begin{eqnarray}
		%     &&n_{\rm{s}}= 1-\frac{2(1+\alpha)}{(2+\alpha)N_{\rm{inf}}}\\
		%     &&r=8 (B\alpha)^{\frac{2}{2+\alpha}}((2+\alpha)N_{\rm{inf}})^{-\frac{2(1+\alpha)}{2+\alpha}}\\
		%     % && \mathcal{P}_{\rm{s}}=A (B\alpha)^{-2+\frac{2(1+\alpha)}{2+\alpha}}((2^{-\frac{2+\alpha}{2}}B\alpha)^{\frac{1}{1+\alpha}}+(2+\alpha)N_{\rm{inf}})^{\frac{2(1+\alpha)}{2+\alpha}} \nonumber\\
		%     &&\mathcal{P}_{\rm{s}}=\frac{A}{12\pi^2} (B\alpha)^{-\frac{2}{2+\alpha}}((2+\alpha)N_{\rm{inf}})^{\frac{2(1+\alpha)}{2+\alpha}}
		% \end{eqnarray}
	% }
% Using the observational constraint on spectral index $n_{\rm{s}}$ for $50<N<60$ and for tensor to scalar ratio $r<0.028$ (BICEP/Keck \cite{BICEP:2021xfz, Galloni:2022mok} and PLANCK \cite{Planck:2018jri}), the constraint on the model parameters can be obtained as:
% \begin{eqnarray}
	% &11.70<\alpha<55.14,\quad B<1.32\times10^{18},&\nonumber\\&1.39 \times 10^{-13}<A<5.91\times10^{-10}&,
	% \end{eqnarray}
%where we use the fact that $B \phi^{-\alpha} \ll 1$, during slow-roll.
In this case, one can relate the model parameter $\alpha$ to the exponent $n$, given in  Eq. \eqref{eq:gen_thetadot-as-theta} as
\begin{eqnarray}\label{eq:small-field-1-n}
	n = \frac{3+2\alpha}{1+\alpha}.
\end{eqnarray}
It is now obvious that, $\alpha > 0$, which in turn, implies:
\begin{eqnarray}
	2 \leq n \leq 3.
\end{eqnarray}
Similarly, $\mu$ and $\nu$ can also be expressed in terms of the models parameters $A$ and $B$ as
\begin{eqnarray}\label{eq:small-field-1-munu}
	\mu = \frac{1}{3}\sqrt{\frac{A}{2}}B^2\alpha^2(1+\alpha)\left(\frac{\alpha B}{\sqrt{6}}\right)^{-\frac{3+2\alpha}{1+\alpha}}, \qquad\qquad\nu =   m.~~
\end{eqnarray}
% {$1.34\times  10^{-5}<\mu <1.44\times 10^{-5}$}\\
Using the above forms of $n, \mu$ and $\nu,$ along with Eqs. \eqref{eq:gen-theta-sol-t}, \eqref{eq:gen-H-theta}, \eqref{eq:gen-eps1-theta}, \eqref{eq:gen-eps2-theta}, \eqref{eq:gen-phidot-theta}, and \eqref{eq:gen-pot-theta}, we can obtain the full solution of the dynamics using our proposed method. However, two other pieces of information are needed to fully solve these equations. The first one is how $\theta_{i},$ i.e., the initial condition for $\theta$ depends on the initial condition of the field $\phi_{i}$, such that Eq. \eqref{eq:gen-theta-sol-t} can be properly solved. This can be obtained by using Eq. \eqref{eq:slow_roll_theta} as:

\begin{eqnarray}\label{eq:small-field-1-thetai}
	\theta_{i}= \frac{\alpha B}{\sqrt{6}}\frac{1}{\phi_{i}^{1 + \alpha}}.
\end{eqnarray}
Note that $\phi_{i}> \phi_*$, where $\phi_*$ relates to the pivot scale $k = 0.05\ \text{Mpc}^{-1}.$ The other information needed is the constant appearing in the general solution of the Hubble parameter in Eq. \eqref{eq:gen-H-theta}. Since the model is categorized under small field models, as mentioned earlier, $C_{\rm{2}}$ can then be expressed as

\begin{eqnarray}\label{eq:small-field-1-c2}
	C_{\rm{2}} = \sqrt{\frac{3}{A}}.
\end{eqnarray}
% {$7.13\times  10^{4}<C_{\rm{2}} <4.64\times 10^{6}$}\\
{ The evolution corresponding to above model for a specific choice of $A$, $B$, and $\alpha$ can be seen in Fig. \ref{fig:poly-n-alpha-plot} and \ref{fig:poly_H_eps_phi_dphi_vs_t} for the domain $10^{-4} \leq \theta \leq \pi/2,$ i.e., from the deep slow-roll inflationary domain to the bottom of the potential. In these figures, we compare our analysis with the numerical solution. As can be seen from these plots, our analysis, similar to chaotic inflation, provides excellent accuracy.\footnote{Please note that, one can see an abrupt yet small deviation in $\epsilon_2$ between our analysis (blue) and the numerical one (with a hinge). This is becuase, while obtaining the numerical solution, we use the two asymptotic solutions given in \eqref{eq:pot-poly} and its derivative to be continuous at the junction. As a consequence, the second-order derivative of the potential at the junction becomes discontinuous, which reflects in the solution of $\epsilon_2.$ However, by closely looking into it, it can also be seen that, except very close to the junction point, both analytical and numerical solutions are in great agreement.}}

%However, at this stage, we must mention that, although the accuracy \viz $\Delta H/H$ varies with the change of the model parmeters, it is still within the same order. For example, below, in Fig. \ref{fig: eff_pot_poly_alp_12}, we show similar analysis for $\alpha = 12$ and in this case, maximum error is $\sim 70 \%$, and at the end of inflation, it is $\sim 38 \%$, which implies that our analysis still provides a great accuracy.}

%
%\begin{figure}
%	\includegraphics[width=0.495\textwidth]{theta_power.pdf}
%	\includegraphics[width=0.495\textwidth]{h_power.pdf}
%	\caption{First kind of small field model: We plot of the evolution of $\theta$ as function of cosmic time $t$ (top) and Hubble parameter $H$ as function of $\theta$ (bottom) with the chosen values of the parameters: $A=10^{-10}$, $m=10^{-5}$, and $\alpha=12$.
%	\label{fig:poly-n-alpha-plot}}
%\end{figure}

%
%\begin{figure}
%	\centering
%	\includegraphics[width=0.8\textwidth]{relative_diff_poly.pdf}
%	\caption{First kind of small field model: we plot the relative difference between the numerical and analytical (proposed solution) of Hubble parameter $H$ as a function of cosmic time $t$ and show that the analytical solution provides a good level of accuracy of evaluating the background dynamics. Here black (dashed) line corresponds to epoch of end of inflation.
%		\label{fig: relative_diff_poly}}
%	\vspace{-5mm}
%\end{figure}

% \begin{eqnarray}
	%     &&\dot{\theta}\simeq\left\{\begin{aligned}
		%     &\mu\theta^\frac{3+2\alpha}{1+\alpha}  \hspace{20pt} \text{Slow-roll} \\ 
		%     &{\nu}  \hspace{45pt} \text{Reheating} \\ 
		%     \end{aligned}\right.,\\
	%     &&\xi=\frac{1}{3}\sqrt{\frac{A}{2}}B^2\alpha^2(1+\alpha)\left(\frac{\alpha B}{\sqrt{6}}\right)^{-\frac{3+2\alpha}{1+\alpha}}.
	% \end{eqnarray}
% \begin{eqnarray}
	% &&\mu= \xi, \quad\nu=m\\
	% &&\quad\theta_{i}= \frac{\alpha B}{\sqrt{6}}\phi_{i}^{-1-\alpha}
	% \end{eqnarray}
% \begin{eqnarray}
	% \dot{\theta}\propto \theta^\frac{3+2n}{1+n}.
	% \end{eqnarray}
\subsection{Second kind of small field inflationary models}

\begin{figure}[h]
	\includegraphics[width=0.495\textwidth]{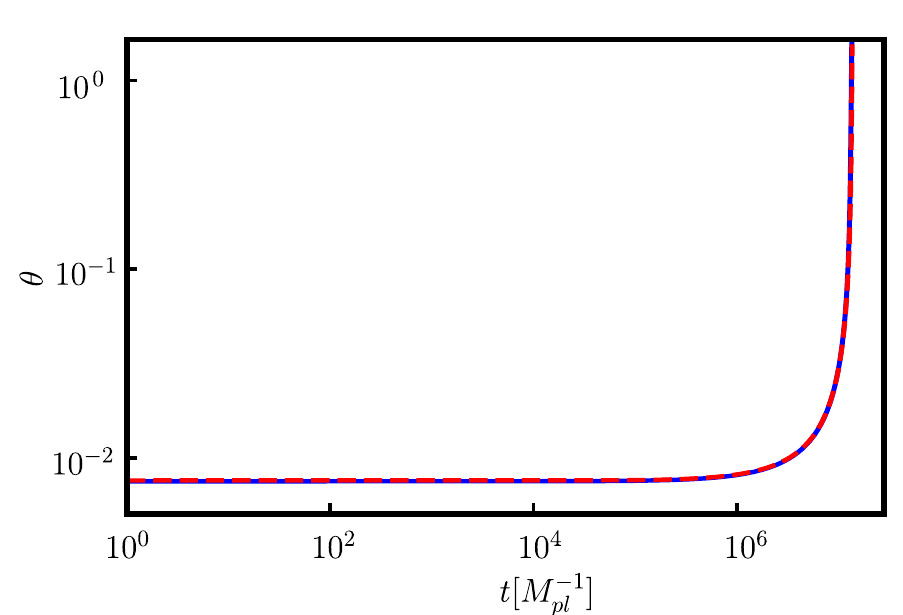}
	\includegraphics[width=0.495\textwidth]{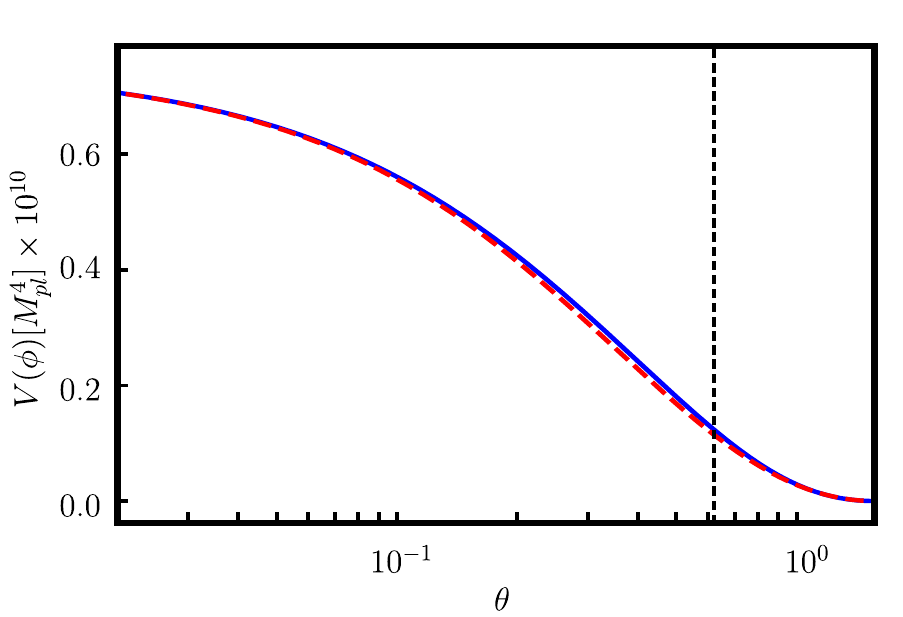}
	
	\caption{Second kind of small field model: We plot the evolution of $ \theta$ as a function of cosmic time $t$ (left) and potential $V(\phi)$ as a function of $\theta$ (right) for $A=3/4 m^2$, $m=10^{-5}$, $B=1$,  and $\alpha=\sqrt{2/3}$, both numerically (dotted-red) and analytically (solid-blue) and show that the analytical solution provides a good level of accuracy of evaluating the background dynamics. The model represents the well-known Starobinsky model of inflation.
		\label{fig:staro_H_vs_theta}}
\end{figure}

\begin{figure}[h]
	\includegraphics[width=0.495\textwidth]{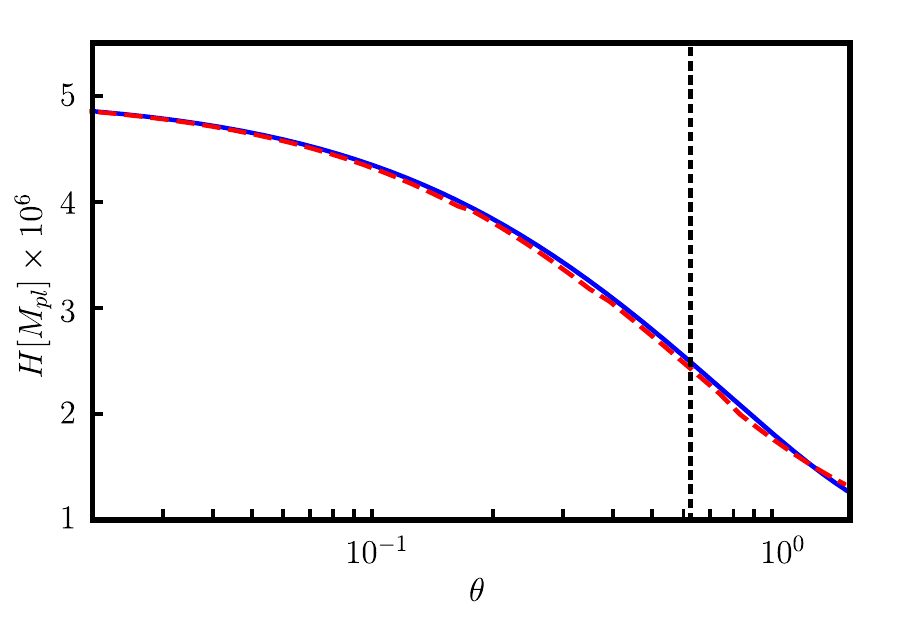}
	\includegraphics[width=0.495\textwidth]{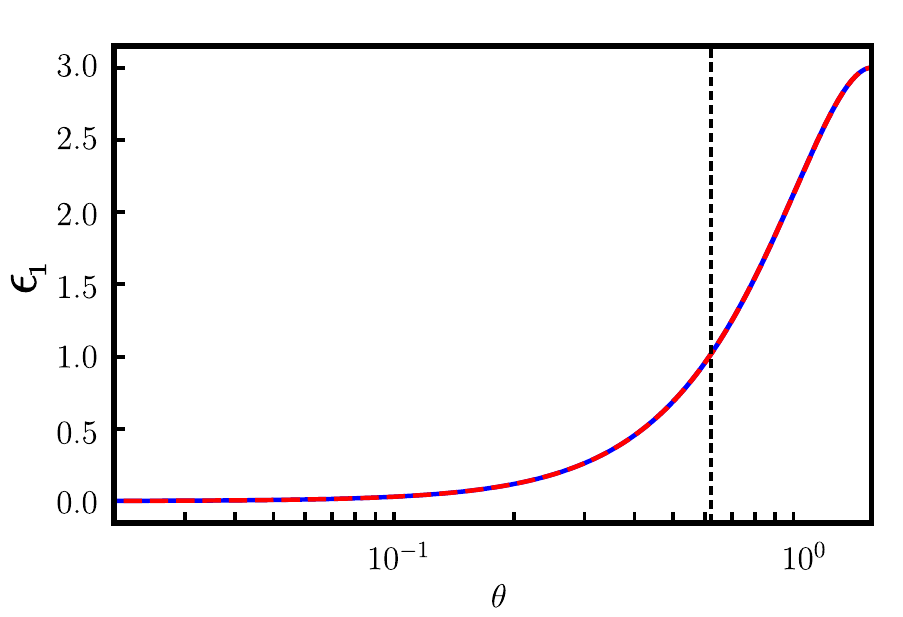}
	\includegraphics[width=0.495\textwidth]{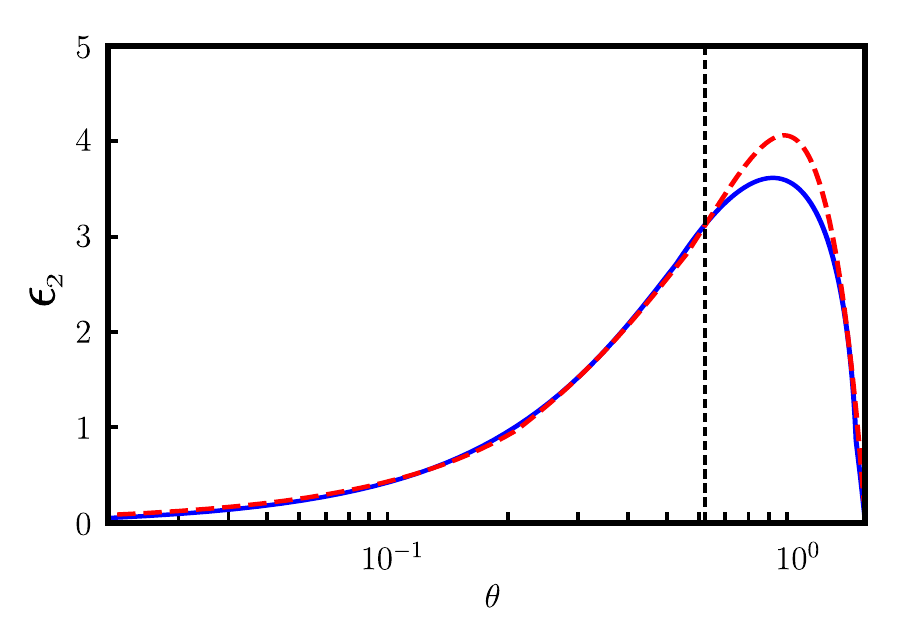}
	\includegraphics[width=0.495\textwidth]{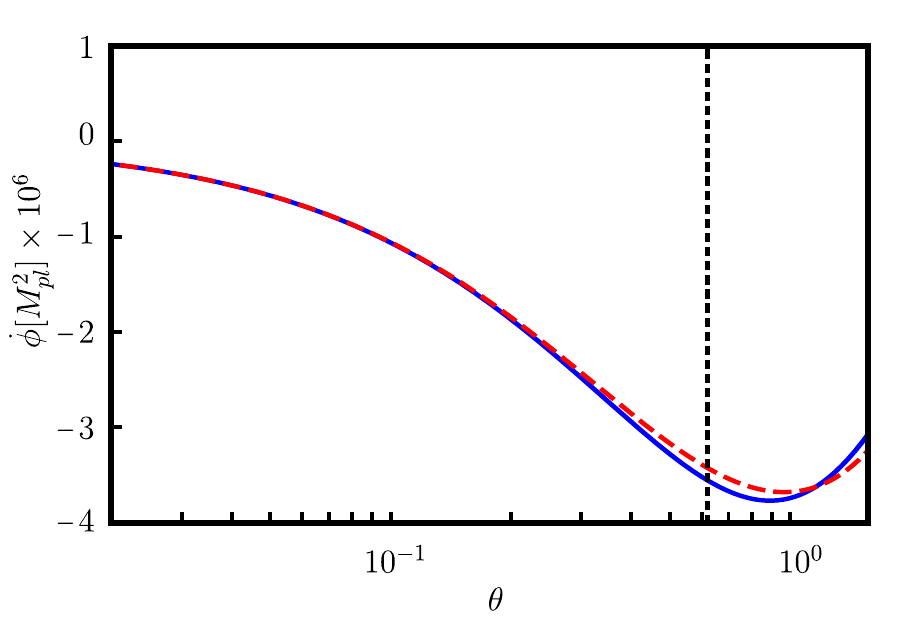}
	\caption{Second kind of small field model: We plot the Hubble parameter $H$ (top-left), first slow-roll parameter $\epsilon_{\rm{1}}$ (top-right), $\phi$ (bottom-left), and $\dot{\phi}$ (bottom-right) as functions of cosmic time $t$ both numerically (dotted-red) and analytically (solid-blue) for $A=3/4 m^2$, $m=10^{-5}$, $B=1$, and $\alpha=\sqrt{2/3}$ and show that the analytical solution provides a good level of accuracy of evaluating the background dynamics.
		\label{fig:staro_H_eps_phi_dphi_vs_t}
		\vspace{-3mm}
	}
\end{figure}

In this case, the potential can be expressed as:
\begin{eqnarray}
	\label{eq: power_law_potential}
	V(\phi)\simeq\left\{\begin{array}{cc}{A\left(1-B e^{-\alpha \phi}\right)} & {~~~ \text{Slow-roll}} \\ { \frac{1}{2}m^2\phi^2} & {~~~~ \text{Reheating}} \\ \end{array}\right.
\end{eqnarray}
where $A$, $B$, $m$, and $\alpha$  are positive constants. This kind of model can be categorized as $\alpha$-attractor model \cite{Kallosh:2022feu, Bhattacharya:2022akq, Carrasco:2015pla, Galante:2014ifa, Kallosh:2013tua, Kallosh:2013yoa}. 
% {
	% \begin{eqnarray}
		%     &&n_{\rm{s}}= 1+\frac{4\alpha^2(-3-\sqrt{2}\alpha-2N_{\rm{inf}}\alpha^2)}{(2+\alpha(\sqrt{2}+2N_{\rm{inf}}\alpha))}\\
		%     &&
		%     r=\frac{32 \alpha^2}{(2+\alpha(\sqrt{2}+2N_{\rm{inf}}\alpha))} \\
		%     &&\mathcal{P}_{\rm{s}}=\frac{A(2+\alpha(\sqrt{2}+2N_{\rm{inf}}\alpha))^2}{48\pi^2\alpha^2}
		% \end{eqnarray}
	% }
% {
	% \begin{eqnarray}
		%     &&n_{\rm{s}}= 1-\frac{2}{N_{\rm{inf}}}\\
		%     &&
		%     r=\frac{8}{N_{\rm{inf}}^2\alpha^2} \\
		%     &&\mathcal{P}_{\rm{s}}=\frac{A N_{\rm{inf}}^2\alpha^2}{12\pi^2}
		% \end{eqnarray}
	% }
% Using the observational constraint on spectral index $n_{\rm{s}}$ for $50<N<60$ and for tensor to scalar ratio $r<0.028$, the constraint on the constants can be given by
% % \begin{eqnarray}
	% % &0.42<\alpha<1.24,\quad B>\frac{\sqrt{2}}{\sqrt{2}+\alpha}&\nonumber\\
	% % &5.1\times10^{-11}<A<3.69\times10^{-10}&
	% % \end{eqnarray}
% \begin{eqnarray}
	% &\alpha > 0.34,\quad B>\frac{\sqrt{2}}{\sqrt{2}+\alpha}&\nonumber\\
	% &A<6.33\times10^{-10}&
	% \end{eqnarray}
where, again, we use the $B e^{-\alpha \phi} \ll 1.$ Similar to earlier case, $n, \mu, \nu, \theta_{i}$ and $C_{\rm{2}}$ can be expressed in terms of the model parameters as

\begin{eqnarray}
	n = 2,\quad \mu= \alpha \sqrt{2}\sqrt{A},\quad\nu=m,\quad \theta_{i}= \frac{\alpha\exp{(-\alpha \phi_{i})}}{\sqrt{6}}, \quad C_{\rm{2}} = \sqrt{\frac{3}{A}}.
\end{eqnarray}
% {$1.17\times  10^{-5}<\mu <1.41\times 10^{-5}$}\\
% {$C_{\rm{2}} >6.88\times 10^{4}$}\\
One example of such a model is the famous Starobinsky model of inflation with the potential given by
% \begin{eqnarray}
	%     \dot{\theta}\simeq\left\{\begin{array}{cc}{\alpha \sqrt{2}\sqrt{A}\ \theta^2} & {~~~ \text{Slow-roll}} \\ { m} & {~~~~ \text{Reheating}} \\ \end{array}\right.
	% \end{eqnarray}
% \begin{eqnarray}
	% \dot{\theta}\propto \theta^2
	% \end{eqnarray}

%%%%%%%%%%%%%%%%%%%%%%%%%%%%%%%%%%%%%%%%%%%
%%%%%%%%%%%%%%%%%%%%%%%%%%%%%%%%%%%%%%%%%%%
%%%%%%%%%%%%%%%%%%%%%%%%%%%%%%%%%%%%%%%%%%%
\begin{eqnarray}\label{eq:Starobinsky potentiall}
	V(\phi)= \frac{3}{4}m^2 \left(1-e^{-\sqrt{\frac{2}{3}}\phi}\right)^2 .
\end{eqnarray}
The corresponding model parameters are thus given by $A=3/4 m^2$, $B=1$, and $\alpha=\sqrt{2/3}$. The evolution corresponding to above example can be seen in Figs. \ref{fig:staro_H_vs_theta} and \ref{fig:staro_H_eps_phi_dphi_vs_t} { and we compare our analytical solution with the numerical ones. As is obvious from these plots, one can see that our analysis is in great agreement with the numerical one\footnote{Note that, unlike the first kind of small field inflationary model given in Eq. \eqref{eq:pot-poly}, in the case of second kind of small field inflationary model given in Eq. \eqref{eq: power_law_potential}, no such abrupt deviation can be seen in $\epsilon_2$. This is because, as for this case, we choose starobinsky model as an example and the corresponding potential and its all higher-order derivatives are continuous at all points. This, in turn, leads to smooth numerical solution.}.}

% For Starobinsky $\mu=\nu=m$, substituting this in \eqref{eq:general_profile} we get,
% \begin{eqnarray}
	% \dot{\theta}&=& \frac{m \theta^2}{1+ \theta^2}\\
	% \theta&=& \frac{1}{2} \left(m (t-t_{i})+\sqrt{4+(-m (t-t_{i})-C_{\rm{1}}){}^2}+C_{\rm{1}}\right)\nonumber\\\\
	% % \end{eqnarray}
% % % \begin{widetext}
	% % \begin{eqnarray}
		%  H&=& (4 \theta  m)\ \left(3 \left(2 \theta ^2-\theta  \sin (2 \theta )+2 \cos (2 \theta )\ +\right.\right.\nonumber\\&&\quad\left.\left.4 \theta\  \text{Si}(2 \theta )-2\right)\ +4 \theta  m~C_{\rm{2}}\right )^{-1}
		% \end{eqnarray}
	% where, $C_{\rm{1}}=\frac{-1+\theta_{i}^2}{\theta_{i}}$, and $C_{\rm{2}} = \frac{2}{m}$.
	% % \end{widetext}
% Now having the above equation we know the form of $\phi,\ \dot{\phi} ~\text{and}~ \epsilon_{\rm{1}}  $.

\section{Observations}\label{sec:obs}

Let us know briefly discuss the observational consequences of these models. During slow-roll, i.e., $\theta \ll 1$, Eqs. \eqref{eq:gen_thetadot-as-theta} and \eqref{eq:gen-Hubble-slow-roll-theta} leads to
\begin{eqnarray}\label{eq:theta_N_slow_roll}
	&&\theta_{N} \equiv \frac{\dot{\theta}}{H} \simeq \mu \theta^n\left(C_{\rm{2}} + \frac{3}{(3 - n)\mu}\ \theta^{3 - n}\right).
\end{eqnarray}
Here, $N \equiv \ln a(t)$ is the e-folding number. Please note that, in the case of large field models, $C_{\rm{2}} = 0,$ whereas, for small field models, $C_{\rm{2}}$ depends on the potential, as mentioned in the Eq. \eqref{eq:C2}. Therefore, during slow-roll, Eq. \eqref{eq:theta_N_slow_roll} can be re-written as
\begin{eqnarray}
	\theta_{N} \simeq \left\{
	\begin{aligned}
		& \frac{3}{3 - n} \theta^3, \quad \text{Large fields,}\\
		& C_{\rm{2}} \mu \theta^n. \quad \quad \text{Small fields}.
	\end{aligned}
	\right.
\end{eqnarray}
For these two separate cases, one can integrate the above equation, and also by using the approximation: $\theta \ll \theta_{\rm end},$ one can then obtain the relation between $\theta$ and $N$ as

\begin{eqnarray}\label{eq:theta-N-slow-roll}
	\theta \simeq \left\{\begin{aligned}
		& \sqrt{\frac{3-n}{6 N}},& \quad \text{Large fields,}\\
		& \left(\frac{1}{C_{\rm{2}} \mu N}\right)^{\frac{1}{n - 1}},& \hspace{13pt} \text{Small fields.}
	\end{aligned}\right.
\end{eqnarray}
% \begin{eqnarray}
	%     &&\theta_*\simeq\frac{1}{\left(C_{\rm{2}} \mu\ N_{\rm{inf}}\right)^{\frac{1}{n-1}}}
	% \end{eqnarray}
Here $N$ denotes the e-folding number from $\theta$ to $\theta_{\rm end}.$ This relation is needed to evaluate the perturbations for a specific $k$ mode. The perturbations, observationally, can be characterized by mainly four parameters: the scalar spectral index $n_{\rm{s}}$, the tensor spectral index $n_{\rm{t}}$, tensor-to-scalar ratio $r$ and the scalar power spectrum $\mathcal{P}_{\rm{s}}.$ For a single canonical scalar field minimally coupled to gravity that leads to slow-roll inflation, these parameters can be written in terms of potential, and the slow-roll parameters as
\begin{eqnarray}
	&&n_{\rm{s}} \simeq 1-2 \epsilon_{\rm{1}}-\epsilon_{\rm{2}}, \qquad\qquad n_{\rm{t}} \simeq - 2 \epsilon_{\rm{1}}, \\
	&&\mathcal{P}_{\rm{s}} \simeq \frac{H^2}{8\pi^2 \epsilon_{\rm{1}}}, \quad\qquad~\qquad\quad\ r \simeq 16 \epsilon_{\rm{1}}.
\end{eqnarray}
Using Eqs. \eqref{eq:gen-Hubble-slow-roll-theta}, \eqref{eq:gen-eps1-slow-roll-theta} and \eqref{eq:gen-eps2-slow-roll-theta}, the above parameters can be expressed in terms of $\theta$ as 

\begin{eqnarray}\label{eq:obs-in-theta}
	&&n_{\rm{s}} \simeq 1+\frac{6(4-n)\theta^2}{n-3}-2 C_{\rm{2}} \mu \theta^{n-1},\ \qquad\qquad n_{\rm{t}} \simeq - 6 \theta^2,\nonumber\\
	&&\mathcal{P}_{\rm{s}} \simeq \frac{1}{24\pi^2\theta^2\left(C_{\rm{2}} + \frac{3}{(3 - n)\mu}\ \theta^{3 - n}\right)^2}, \qquad\qquad r \simeq 48\theta^2.
\end{eqnarray}

Observations (BICEP/Keck \cite{BICEP:2021xfz,Galloni:2022mok} and PLANCK \cite{Planck:2018jri}) suggest that, at the pivot scale ($k_*=0.05\, \text{Mpc}{}^{-1}$), the amplitude of the scalar power spectrum is $\mathcal{P}_{\rm{s}} \simeq 2.101^{+0.031}_{-0.034} \times 10^{-9} (68\%\, \, \text{CL})$ with the scalar spectral index being $n_{\rm{s}} = 0.9649 \pm 0.0042\,(68\%\, \, \text{CL})$, while the tensor-to-scalar ratio $r$ is bounded from above by $r<0.028\, (95 \%\, \, \text{CL})$. As of yet, there is no bound on the tensor spectral index $n_{\rm{t}}.$ To evaluate Eq. \eqref{eq:obs-in-theta}, an additional requirement is necesaary: the relation between $\theta$ related to the pivot scale, i.e., $\theta_*$ and the e-folding number $N,$ given in Eq. \eqref{eq:theta-N-slow-roll}. In general, the pivot scale leaves the Hubble horizon $50 - 60$ e-folds before the end of inflation, i.e., $N_* \sim 50 - 60.$ Therefore, by using this information, one can evaluate the observational parameters for any model of inflation.

Let us first discuss the large field models with $V(\phi) \propto \phi^\alpha,~\alpha > 0$.  In this case,   one can verify that
\begin{eqnarray}
	n = 3 - \frac{\alpha}{2}.
\end{eqnarray}
It leads to $r  = 4\alpha/N_{*}$ and $n_{\rm{s}} = 1 - (2 + \alpha)/(2 N_{*}),$ which, one can then quickly verify that for $N_* \sim 50 - 60,$ that these relations do not obey the observational constraints, as is already mentioned in the previous section. On the other hand, in the case of small field models, all observational parameters can be expressed in terms of $C_{\rm{2}}, \mu$ and $n$ as
%\textbf{Small Field models}
% {
	% \begin{eqnarray}
		%     % &&n_{\rm{s}}= 1-\frac{2(1+\alpha)}{(2+\alpha)N_{\rm{inf}}}\\
		%     &&n_{\rm{s}}=1-2 ((n-1)N_{\rm{inf}})^{-1}-\frac{6(4-n)}{3-n(C_{\rm{2}} \mu (n-1)N_{\rm{inf}})^{-\frac{2}{n-1}}}\nonumber\\
		%     &&
		%     r=48 (C_{\rm{2}} \mu (n-1)N_{\rm{inf}})^{-\frac{2}{n-1}}\\
		%     &&\mathcal{P}_{\rm{s}}=\frac{(C_{\rm{2}} \mu (n-1)N_{\rm{inf}})^{\frac{2}{n-1}}}{24\pi^2\left(C2+\frac{3}{(3-n)\mu}(C_{\rm{2}} \mu (n-1)N_{\rm{inf}})^{-\frac{3-n}{n-1}}\right)^2}
		% \end{eqnarray}
	% }
\begin{eqnarray}
	% &&n_{\rm{s}}= 1-\frac{2(1+\alpha)}{(2+\alpha)N_{\rm{inf}}}\\
	& n_{\rm{s}} \simeq& 1 - \frac{2}{(n-1)N_*},\\
	&n_{\rm{t}} \simeq& -\frac{6}{(C_{\rm{2}} \mu (n-1)N_*)^{\frac{2}{n-1}}},\\
	&r \simeq&\frac{48}{(C_{\rm{2}} \mu (n-1)N_*)^{\frac{2}{n-1}}},\\
	&\mathcal{P}_{\rm{s}} \simeq&\frac{(C_{\rm{2}} \mu (n-1)N_*)^{\frac{2}{n-1}}}{24\pi^2 C_{\rm{2}}^2}.
\end{eqnarray}
Again, by using the observational constraints with $N_* \sim 50 - 60$, one can, in general, obtain the constrained values of these parameters as

\begin{eqnarray}
	1.84<n<2.29,\quad C_{\rm{2}}>5.87\times 10^{4},\quad
	0.83\times  10^{-5}<\mu <3.21\times 10^{-5}.
\end{eqnarray}
These are the most general constraints on small-field inflationary models.

Let us now focus separately on the two different and special cases of small field models that we discussed in the previous section. For the first kind of small field models with $V(\phi)=A(1-B\phi^{-\alpha})$, all observational parameters can be expressed as

\begin{eqnarray}
	&n_{\rm{s}} \simeq& 1-\frac{2(1+\alpha)}{(2+\alpha)N_*},\\
	&n_{\rm{t}} \simeq& - \frac{(B\alpha)^{\frac{2}{2+\alpha}}}{((2+\alpha)N_*)^{\frac{2(1+\alpha)}{2+\alpha}}},\\
	&r \simeq&\frac{8 (B\alpha)^{\frac{2}{2+\alpha}}}{((2+\alpha)N_*)^{\frac{2(1+\alpha)}{2+\alpha}}},\\
	% && \mathcal{P}_{\rm{s}}=A (B\alpha)^{-2+\frac{2(1+\alpha)}{2+\alpha}}((2^{-\frac{2+\alpha}{2}}B\alpha)^{\frac{1}{1+\alpha}}+(2+\alpha)N_{\rm{inf}})^{\frac{2(1+\alpha)}{2+\alpha}} \nonumber\\
	&\mathcal{P}_{\rm{s}} \simeq&\frac{A}{12\pi^2} \frac{((2+\alpha)N_*)^{\frac{2(1+\alpha)}{2+\alpha}}}{(B\alpha)^{\frac{2}{2+\alpha}}}.
\end{eqnarray}
Please note that these expressions are obtained by using Eqs. \eqref{eq:theta-N-slow-roll} and \eqref{eq:obs-in-theta}, which, one can verify to be true by using $\phi$ and $N$ relations, as used in general. It, therefore, shows that our method also provides consistent results for the perturbations. Using these observational constraints, the constraint on the model parameters, for this special case, can be obtained as:
\begin{eqnarray}
	2.40\leq\alpha\leq 55.14, \qquad A \leq 8.70\times10^{-10},
\end{eqnarray}
% \beqObservations (BICEP/Keck \cite{BICEP:2021xfz} and PLANCK \cite{Planck:2018jri}), suggest that, at the pivot scale ($k_*=0.05\, \text{Mpc}{}^{-1}$), the amplitude of the scalar power spectrum is $P_{\rm{s}} \simeq 2.101^{+0.031}_{-0.034} \times 10^{-9} (68\%\, \, \text{CL})$ with the scalar spectral index being $n_{\rm{s}} = 0.9649 \pm 0.0042\,(68\%\, \, \text{CL})$, while the ratio $(r)$ of the tensor-to-scalar perturbation amplitudes is bounded by $r<0.036\, (95 \%\, \, \text{CL})$.
% 1.33\times  10^{-5}<\mu <1.48\times 10^{-5}
% \eeq
and $B$ can be as large as $10^{126}$.
%\begin{eqnarray}
%&1.34\times  10^{-5}<\mu <1.44\times 10^{-5}&\\
%&7.13\times  10^{4}<C_{\rm{2}} <4.64\times 10^{6}&
%\end{eqnarray}

In the second special case with $V(\phi)=A(1-B\exp(-\alpha \phi))$, similarly, the observable parameters can be written in terms of the e-folding number as

\begin{eqnarray}\label{eq:staro_ns}
	&n_{\rm{s}} \simeq & 1-\frac{2}{N_*},\\
	&n_{\rm{t}} \simeq& - \frac{1}{\alpha^2 N_*^2}, \\
	&r \simeq &\frac{8}{\alpha^2N_*^2}, \\
	&\mathcal{P}_{\rm{s}} \simeq&\frac{A \alpha^2 N_*^2}{12\pi^2}.
\end{eqnarray}

Using the observational constraints, the constraint on the model parameters can be obtained as
% \begin{eqnarray}
	% &0.42<\alpha<1.24,\quad B>\frac{\sqrt{2}}{\sqrt{2}+\alpha}&\nonumber\\
	% &5.1\times10^{-11}<A<3.69\times10^{-10}&
	% \end{eqnarray}

\begin{eqnarray}
	\alpha \geq 0.34,\qquad A\leq6.33\times10^{-10}.
\end{eqnarray}

It is important to note that the constraints on the model parameters mentioned above do not account for the effect of reheating; therefore they do not justify the actual limitations. To include the effect, we must analyze the effective equation of state parameter during reheating $w_{\rm{re}}$, and the duration of reheating $N_{\rm{re}}$, which is governed by the equation \cite{Liddle:2003as}: 

\begin{eqnarray}\label{eq:Nre}
	&& \hspace{-10pt} N_{\rm{re}} = \frac{4}{3w_{\rm{re}}-1}\left(\log\left(
	\frac{k}{a_{\rm{0}} T_{\rm{0}}}\right)+N_k-\log(H_k)+\right.\nonumber\\
	&& \hspace{-10pt}\left. \frac{1}{4}\log(\rho_{\rm{end}})+\frac{1}{3}\log\left(\frac{11 g_{\rm{s,re}}}{43}\right)+\frac{1}{4}\log\left(\frac{30}{\pi^2 g_{\rm{reh}}}\right)\right). \qquad
\end{eqnarray}

Here, $\{a_{\rm{0}}, T_{\rm{0}}\}$ are present values of the scale factor and the temperature of the Universe, respectively. $H_k$ is the Hubble scale when the mode leaves the Horizon; $\rho_{\rm{end}}$ is the energy density at the end of inflation and $\{g_{\rm{reh}}, g_{\rm{s, re}}\}$ are the effective number of relativistic species upon thermalization and the effective number of light species for entropy during reheating, respectively.  The result obtained from this kind of simplified yet significant study is known as the quantitative analysis of reheating, and it has been the subject of extensive academic study \cite{Martin:2006rs, Lorenz:2007ze, Martin:2010kz, Adshead:2010mc, Mielczarek:2010ag, Easther:2011yq, Dai:2014jja, Domcke:2015iaa, Lozanov:2016hid, Kabir:2016kdh, Liddle:2003as, Martin:2014nya, Maity:2016uyn, Maity:2018qhi, Maity:2019ltu, Nandi:2019xve,Turner:1983he, Martin:2013tda, Dai:2014jja}. However, since these studies utilize simplified slow-roll and reheating approximations, the analysis can be significantly enhanced with the use of our proposed method on accurately representing inflationary dynamics. Consequently, to determine the overall contribution of all the factors listed on the right-hand side of the above equation, it is necessary to meticulously incorporate the accurate dynamics of the inflationary paradigm for each term, specifically the second, third, and fourth terms from the aforementioned equation\footnote{One, in principle, should also be careful in interpreting the constant effective equation of state parameter $w_{\rm re}.$}.
\begin{figure}[h]
	\centering
	\includegraphics[width=0.8\textwidth]{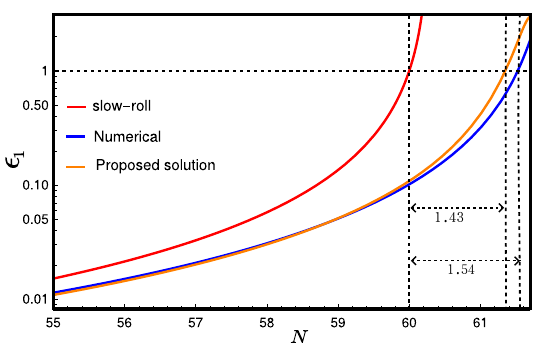}
	\caption{Starobinsky inflation: We plot the first slow roll parameter $\epsilon_{\rm{1}}$ as functions of e-fold variable $N$ numerically (blue), slow-roll approximation (red), and our proposed analytical solution (orange) with the black (dashed) lines corresponds to epoch end of inflation for each curve. It is easy to understand that the end of inflation happens at different values of $N$, even if one starts with idetical initial conditions.}
	\label{fig: deltaN_comparison}
\end{figure}

Let's now examine the most straightforward but effective corrections obtained from the first term on the right side: $N_k$, in order to quickly comprehend the implications of it. In Fig. \ref{fig: deltaN_comparison}, we plot the first slow-roll parameter for the Staobinsky model of inflation using the leading order slow-roll approximation and numerical analysis (and our method). It is easy to understand that the end of inflation occurs at different values of $N$ for both of these techniques, even if one starts with idetical initial conditions. For convenience sake, let's choose $\phi_i = 5.4315,~\phi_{,N} = - V'(\phi_i)/V(\phi_i) = -0.0192$, for instance. Using leading order slow-roll approximation, this results in 60 e-folds of inflation, and the equivalent value of $\phi_{\rm end}$ is 0.9402. Numerical simulation (and our method) result in 61.5237 (61.4388) e-folds of inflation with $\phi_{\rm end} \simeq 0.6146$ under the same conditions. The numerical difference $\Delta N$ is roughly 1.5 e-folds (as in our case, $\Delta N \sim 0.1$). For the sake of simplicity, let us suppose that late time physics with instantaneous reheating (i.e., Eq. \eqref{eq:Nre} with $N_{\rm re} = 0$)\footnote{In fact, many models of reheating now prefers instanteneous reheating. However, the upper limit on $N_{\rm re}$ is achieved such that at the end of reheating, the energy is higher than the electroweak scale $\sim 150$ GeV. Sometimes, a stronger scale is considered as the BBN scale of $\sim 1 $MeV.} fixes the value $N_k$ at 55 e-folds (before the end of inflation). In that instance, the numerical simulation yields the value of scalar field at pivot scale: $\phi_k \simeq 5.3213$, whereas, the leading order slow-roll approximation yields  $\phi_k \simeq 5.3529$. This discrepancy arises from the difference in the end of inflation rather than the slow-roll approximation. Consequently, numerical simulation results in $\epsilon_1 \simeq 2.2769 \times 10^{-4},~\epsilon_2 \simeq 3.5099 \times 10^{-2}$ at the pivot scale, where the slow-roll approximation yields $\epsilon_1 \simeq 2.1864 \times 10^{-4},~\epsilon_2 \simeq 3.4586 \times 10^{-2}$. Using the leading order slow-roll approximation, the scalar spectral index for Starobisky inflation can be expressed as (see, for instance, Eq. \eqref{eq:staro_ns})\footnote{Please note that, this equation itself is the result of slow-roll approximation and  may significantly be improved if we incorporate accurate inflationary dynamics.}
\begin{eqnarray}
	n_s \simeq  1 - 2 \epsilon_1 - \epsilon_2,
\end{eqnarray}
evaluated at the pivot scale. Thus, for the Starobinsky inflation, the scalar spectral index at the pivot scale, determined by the slow-roll approximation, is $n_s \simeq 0.965$, whereas the numerical simulation yields $n_s \simeq 0.964$. This is expected as, using slow-roll approximation, one can express $\epsilon_1$ and $\epsilon_2$ in terms of the e-folding number which makes
\begin{eqnarray}
	n_s\sim 1-\frac{2}{N_k},\qquad\qquad \Delta n_s\sim\frac{2}{N_k^2}\Delta N_k.\nonumber
\end{eqnarray}
As can be seen from the above calculation, given our situation, where $\Delta N_k \simeq 1.5$, the discrepancy in $n_s$ becomes $\Delta n_s \simeq 10^{-3}$.
Even though this discrepancy might not appear significant right now, instead, it might potentially be seen by future experiments like PRISM \cite{PRISM:2013ybg} and EUCLID \cite{EuclidTheoryWorkingGroup:2012gxx}, cosmic 21-cm surveys \cite{Mao:2008ug}, and CORE experiments \cite{CORE:2016ymi} with increased precision of $10^{-3}$ in $n_s$, which would strongly constrain different inflation models. Similarly, the theoretical predictions can be further improved if all changes pertaining to all the terms in the right-hand side are taken into account in a suitable manner. Since it requires a thorough investigation, we save it for future work.
\section{Conclusions}\label{sec:conclu}
% \begin{itemize}
	%     \item  In this work, we have constructed a single analytical solution during the early Universe for the case of a single canonical scalar field minimally coupled to gravity. 
	%     \item To achieve such we have make use of variables $\{H,\theta\}$ instead of $\{\phi,\dot{\phi}\}$.
	%     \item we know in great detail the solution during the slow roll and the solution during the deep oscillatory phase.
	%     \item The only analytical solution we are unaware of is the smooth transition from the inflationary solution to the oscillatory phase.
	%     \item This solution will provide great insight into the qualitative and quantitative analysis of reheating, which we will study in the future.
	%     \item will provide better constraints on parameters using both analysis.
	%     \item This can improve bounds on ns and r.
	% \end{itemize}
% In this work, we constructed a single analytical solution of a single canonical scalar field minimally coupled to gravity.
% % i.e., Eqs. \eqref{eq:energy-eq}, \eqref{eq:acc-eq} and \eqref{eq:phi} (two of them are independent). 

% This has long been solved for two different regimes: the slow-roll phase and the reheating phase. However, the full solution of the system from slow-roll to smoothly transiting into the reheating oscillatory phase of the inflaton field was unknown. More precisely, the solution around the end of inflation and the beginning of the reheating period was not known. In this work, we propose a method to resolve this problem. In doing so, we assume:\\

In this  article, we considered a single canonical scalar field model minimally coupled to gravity with a potential $V(\phi)$ that leads to the evolution of the Universe consisting of both slow-roll inflation and oscillatory behavior around the potential's minimum, also known as the reheating era. The complete solution for the background field in this scenario remains elusive. Traditionally, we solve it using two discrete regimes, each with its own set of conditions.

The first regime is characterized by the slow-roll condition, in which the slow-roll parameters are considerably less than unity. In this regime, the slow-roll solution is a well-established analytical solution. In contrast, the second regime takes effect when the field enters the reheating phase, and the Hubble parameters fall substantially below the effective mass of the potential.

The difficulty arises, however, when attempting to bridge the distance between these two regimes during a phase in which both slow-rolling and reheating conditions fail. In order to resolve this dilemma, our work seeks to present a model-independent, unified solution. With this objective in mind, we assumed the following:

\begin{enumerate}
	\item {\bf Simple potential:} the nature of the potential is simple.
	\item {\bf Minimum of the potential:} potential has a minimum at $\phi = 0$.
	\item  {\bf Exact de-Sitter:} $|\phi| \rightarrow \infty$ leads to $\epsilon_{\rm{1}} = 0,$ implying the de-Sitter Universe.
	\item {\bf Slow-roll:} the potential, for $|\phi|\gg 1$ leads to slow-roll inflation.
	\item {\bf Near-minimum behavior:} In the vicinity of the minimum, the potential can be approximated as $V(\phi) \propto \phi^2$.
\end{enumerate}

To address the difficulty, we considered a new approach by employing phase-space variables $\{\theta, H\}$ rather than $\{\phi, \dot{\phi}\}$. The advantage of this method is that the variable $\theta$ corresponds directly to various phases of cosmic evolution. For example, $\theta \ll 1$ (implies $|\phi| \gg 1$) denotes a period of slow-roll inflationary evolution, whereas $\theta = \sin ^{-1} (1/\sqrt{3})$ (equivalent to $\epsilon_{\rm{1}} = 1$) denotes the precise end of the inflationary phase. In addition, $\theta = \pi/2$ corresponds to $\phi = 0$, indicating that the field has reached the minimal potential for the first time, $\theta = \pi$ implies the field velocity $\dot{\phi}$ is zero, and so on. It is particularly significant because we know that particle production, specifically resonance, occurs at the potential's minimum --- a location that is difficult to pinpoint using cosmic time, $t$.

In our work, we first propose a unified solution for the Universe within the chaotic inflation model, where the potential is $V(\phi) = 1/2 m^2 \phi^2$. We provide comprehensive solutions for critical parameters such as the Hubble parameter $H$, slow-roll parameters $\epsilon_{\rm{1}}$ and $\epsilon_{\rm{2}}$, the scalar field $\phi$, and its time derivative $\dot{\phi}$ --- all expressed in terms of $\theta$. To complete the dynamics, we also furnish the solution for $\theta$ as a function of cosmic time, $t$. We further extend our methods to broader models of inflation. This accomplishment addresses three essential concerns:
\begin{itemize}
	\item[(a)] {\bf Full Dynamics:} We now possess the complete dynamical evolution of the Universe, spanning from slow-roll inflation to reheating, including the intermediate junction between these phases, rendered smoothly.
	\item[(b)]  {\bf Intermediate identification:} Second, as the full solution is now known, one can immediately identify the evolution simply by knowing the correct value of $\theta$ and analyzing the solution for the complete evolution of the Universe.
	\item[(c)] {\bf Model Independence:} Our solution is not tied to a specific model; instead, it can be applied across a wide spectrum of inflationary models.
\end{itemize} 

This integrated solution, for the case of Chaotic inflation,  has the relative error (compared to the numerical one) that varies from less tha $1 \%$ (deep slow-roll and deep reheating phase) to a peak $\sim 15 \%$ (around the end of inflation), which tells us that the our approximated method provides a good depiction of the inflationary as well as the reheating dynamics. It also provides insightful qualitative and quantitative analysis of reheating. On the qualitative front, we can now incorporate the effects of the end of inflation and the onset of reheating during the creation of particles via parametric resonance, a process effectively described by the Mathieu equation \cite{Amin:2014eta}. On the other hand, as stated previously, for quantitative analysis, the equation that relates the reheating e-folding number $N_{\rm{re}}$ (or the temperature at the end of reheating, i.e., $T_{\rm{re}}$) to the scalar spectral index $n_{\rm{s}}$, i.e., Eq. \eqref{eq:Nre}, can be modified by incorporating the proposed solution, which we believe can help improve the observational constraints. For example, in the case of Starobinsky model, without even considering the complete effect due to it, a simple correction due to the (in)accuracy of the end of inflation, we showed that the scalar spectral index can be improved in the order $10^{-3},$ which using future experiments like PRISM \cite{PRISM:2013ybg} and EUCLID \cite{EuclidTheoryWorkingGroup:2012gxx}, cosmic 21-cm surveys \cite{Mao:2008ug}, and CORE experiments \cite{CORE:2016ymi} can be constrained. Besides, the primordial gravitational waves provide a compelling window to look at physics beyond the BBN and thereby may constrain the reheating phase \cite{Boyle:2007zx, Koh:2018qcy}. In conclusion, by combining these two analysis, the theoretical predictions can be substantially enhanced.
%\cite{Liddle:2003as}: 
%\begin{eqnarray}
%	&& N_{\rm{re}} = \frac{4}{3w_{\rm{re}}-1}\left(\log\left(
%	\frac{k}{a_{\rm{0}} T_{\rm{0}}}\right)+N_k-\log(H_k)+\right.\nonumber\\
%	&&\left. \frac{1}{4}\log(\rho_{\rm{end}})+\frac{1}{3}\log\left(\frac{11 g_{\rm{s,re}}}{43}\right)+\frac{1}{4}\log\left(\frac{30}{\pi^2 g_{\rm{reh}}}\right)\right).~~~~~~~
%\end{eqnarray}

%Here, $\{a_{\rm{0}}, T_{\rm{0}}\}$ are present values of the scale factor and the temperature of the Universe, respectively. $H_k$ is the Hubble scale when the mode leaves the Horizon; $\rho_{\rm{end}}$ is the energy density at the end of inflation and $\{g_{\rm{reh}}, g_{s, re}\}$ are the effective number of relativistic species upon thermalization and the effective number of light species for entropy during reheating, respectively. 
In conclusion, we present a concise summary of the benefits and drawbacks, as well as our interests and our desired future course, which we believe will benefit our community overall. In this study, we presented a new parametrization method-based approximation (using $\theta$), and we demonstrated that the suggested unified solution delivers a new, near-accurate description of both the inflationary as well as reheating solutions. This, in-turn, can be utilized in future experiments to constrain numerous inflationary models \cite{PRISM:2013ybg, EuclidTheoryWorkingGroup:2012gxx, Mao:2008ug, CORE:2016ymi, Boyle:2007zx, Koh:2018qcy}. While we mostly studied small-field models like $\alpha$-attractor models of inflation, we acknowledge that there are more intricate situations. These include models with discrete behaviors near the minimum that do not follow $\phi^2$, as well as models that diverge from slow-roll during inflation, including those that can produce primordial black holes \cite{Garcia-Bellido:2017mdw, Germani:2017bcs, Ballesteros:2017fsr, Dalianis:2018frf}. That being said, it remains a useful method for constraining other existing simpler models. It might be argued that while full numerical simulations for models with a single parameter are feasible, studying multi-parameter models presents a significant difficulty due to their high numerical complexity. When such circumstances arise, our methodology can be readily executed. One can also extend this analysis to check implications on the non-Gaussian observables, such as the non-Gaussianity parameter $f_{\rm NL}$ \cite{Maldacena2003, Chen:2006nt, Nandi:2015ogk, Nandi:2016pfr, Nandi:2017pfw}, in a manner similar to that of gaussian observables (in our case, scalar spectral index $n_s$ or the tensor-to-scalar ratio $r$). The qualitative reheating analysis, which explains the microphysics of reheating, is another topic that we totally overlooked in this work, can also be studied extensively. In recent times, numerous feasible bouncing models \cite{Nandi:2018ooh, Nandi:2019xlj,Nandi:2019xag, Nandi:2020sif, Nandi:2020szp, Nandi:2022twa, Nandi:2023ooo, Kaur:2023uaz} have been developed utilizing the slow-roll inflationary model. These models, once more, can be examined through our approach. Finally, the scope of the suggested method is enormous and it can be improved in many ways. Further exploration of these intricate and numerous topics through comprehensive perturbation analysis is reserved for our future pursuits.
\vspace{-3mm}

\section*{Acknowledgements}
DN is supported by the DST, Government of India through the DST-INSPIRE Faculty fellowship (04/2020/002142). MK is supported by a DST-INSPIRE Fellowship under the reference number: IF170808, DST, Government of India. DN and MK are also very thankful to the Department of Physics and Astrophysics, University of Delhi. MK and DN also acknowledge facilities provided by the IUCAA Centre for Astronomy Research and Development (ICARD), University of Delhi.

\bibliographystyle{JHEP}
%\bibliography{my_collection}

\begin{thebibliography}{10}
	
	\bibitem{STAROBINSKY198099}
	A.~Starobinsky, {\it A new type of isotropic cosmological models without
		singularity},  {\em Physics Letters B} {\bf 91} (1980), no.~1 99 -- 102.
	
	\bibitem{Guth:1981}
	A.~H. Guth, {\it Inflationary universe: A possible solution to the horizon and
		flatness problems},  {\em Phys. Rev. D} {\bf 23} (Jan, 1981) 347--356.
	
	\bibitem{Sato:1981}
	K.~Sato, {\it {First-order phase transition of a vacuum and the expansion of
			the Universe}},  {\em Monthly Notices of the Royal Astronomical Society} {\bf
		195} (07, 1981) 467--479,
	[\href{http://xxx.lanl.gov/abs/http://oup.prod.sis.lan/mnras/article-pdf/195/3/467/4065201/mnras195-0467.pdf}{{\tt
			http://oup.prod.sis.lan/mnras/article-pdf/195/3/467/4065201/mnras195-0467.pdf}}].
	
	\bibitem{Mukhanov:1981xt}
	V.~F. Mukhanov and G.~V. Chibisov, {\it {Quantum Fluctuations and a Nonsingular
			Universe}},  {\em JETP Lett.} {\bf 33} (1981) 532--535. [Pisma Zh. Eksp.
	Teor. Fiz.33,549(1981)].
	
	\bibitem{LINDE1982389}
	A.~Linde, {\it A new inflationary universe scenario: A possible solution of the
		horizon, flatness, homogeneity, isotropy and primordial monopole problems},
	{\em Physics Letters B} {\bf 108} (1982), no.~6 389 -- 393.
	
	\bibitem{HAWKING1982295}
	S.~Hawking, {\it The development of irregularities in a single bubble
		inflationary universe},  {\em Physics Letters B} {\bf 115} (1982), no.~4 295
	-- 297.
	
	\bibitem{STAROBINSKY1982175}
	A.~Starobinsky, {\it Dynamics of phase transition in the new inflationary
		universe scenario and generation of perturbations},  {\em Physics Letters B}
	{\bf 117} (1982), no.~3 175 -- 178.
	
	\bibitem{Guth:1982}
	A.~H. Guth and S.-Y. Pi, {\it Fluctuations in the new inflationary universe},
	{\em Phys. Rev. Lett.} {\bf 49} (Oct, 1982) 1110--1113.
	
	\bibitem{Albrecht-Steinhardt:1982}
	A.~Albrecht and P.~J. Steinhardt, {\it Cosmology for grand unified theories
		with radiatively induced symmetry breaking},  {\em Phys. Rev. Lett.} {\bf 48}
	(Apr, 1982) 1220--1223.
	
	\bibitem{Albrecht:1982mp}
	A.~Albrecht, P.~J. Steinhardt, M.~S. Turner, and F.~Wilczek, {\it {Reheating an
			Inflationary Universe}},  {\em Phys. Rev. Lett.} {\bf 48} (1982) 1437.
	
	\bibitem{Linde:1983gd}
	A.~D. Linde, {\it {Chaotic Inflation}},  {\em Phys. Lett.} {\bf 129B} (1983)
	177--181.
	
	\bibitem{VILENKIN1983527}
	A.~Vilenkin, {\it Quantum fluctuations in the new inflationary universe},  {\em
		Nuclear Physics B} {\bf 226} (1983), no.~2 527 -- 546.
	
	\bibitem{Bardeen:1983}
	J.~M. Bardeen, P.~J. Steinhardt, and M.~S. Turner, {\it Spontaneous creation of
		almost scale-free density perturbations in an inflationary universe},  {\em
		Phys. Rev. D} {\bf 28} (Aug, 1983) 679--693.
	
	\bibitem{1990-Kolb.Turner-Book}
	E.~W. Kolb and M.~Turner, {\em The early universe}.
	\newblock Reading, Mass. : Addison-Wesley, May, 1990.
	
	\bibitem{Mukhanov:1990me}
	V.~F. Mukhanov, H.~A. Feldman, and R.~H. Brandenberger, {\it {Theory of
			cosmological perturbations. Part 1. Classical perturbations. Part 2. Quantum
			theory of perturbations. Part 3. Extensions}},  {\em Phys. Rept.} {\bf 215}
	(1992) 203--333.
	
	\bibitem{Liddle:1994dx}
	A.~R. Liddle, P.~Parsons, and J.~D. Barrow, {\it {Formalizing the slow roll
			approximation in inflation}},  {\em Phys. Rev. D} {\bf 50} (1994) 7222--7232,
	[\href{http://xxx.lanl.gov/abs/astro-ph/9408015}{{\tt astro-ph/9408015}}].
	
	\bibitem{2005hep.th....3203L}
	A.~{Linde}, {\it {Particle Physics and Inflationary Cosmology}},  {\em arXiv
		e-prints} (Mar., 2005) hep--th/0503203,
	[\href{http://xxx.lanl.gov/abs/hep-th/0503203}{{\tt hep-th/0503203}}].
	
	\bibitem{Bassett:2005xm}
	B.~A. Bassett, S.~Tsujikawa, and D.~Wands, {\it {Inflation dynamics and
			reheating}},  {\em Rev. Mod. Phys.} {\bf 78} (2006) 537--589,
	[\href{http://xxx.lanl.gov/abs/astro-ph/0507632}{{\tt astro-ph/0507632}}].
	
	\bibitem{Sriramkumar:2009kg}
	L.~Sriramkumar, {\it {An introduction to inflation and cosmological
			perturbation theory}},  \href{http://xxx.lanl.gov/abs/0904.4584}{{\tt
			arXiv:0904.4584}}.
	
	\bibitem{Baumann:2009ds}
	D.~Baumann, {\it {Inflation}},  in {\em {Physics of the large and the small,
			TASI 09, proceedings of the Theoretical Advanced Study Institute in
			Elementary Particle Physics, Boulder, Colorado, USA, 1-26 June 2009}},
	pp.~523--686, 2011.
	\newblock \href{http://xxx.lanl.gov/abs/0907.5424}{{\tt arXiv:0907.5424}}.
	
	\bibitem{Linde:2014nna}
	A.~Linde, {\it {Inflationary Cosmology after Planck 2013}},  in {\em
		{Proceedings, 100th Les Houches Summer School: Post-Planck Cosmology: Les
			Houches, France, July 8 - August 2, 2013}}, pp.~231--316, 2015.
	\newblock \href{http://xxx.lanl.gov/abs/1402.0526}{{\tt arXiv:1402.0526}}.
	
	\bibitem{Martin:2015dha}
	J.~Martin, {\it {The Observational Status of Cosmic Inflation after Planck}},
	{\em Astrophys. Space Sci. Proc.} {\bf 45} (2016) 41--134,
	[\href{http://xxx.lanl.gov/abs/1502.0573}{{\tt arXiv:1502.0573}}].
	
	\bibitem{Ade:2015lrj}
	{\bf Planck} Collaboration, P.~Ade {\em et.~al.}, {\it {Planck 2015 results.
			XX. Constraints on inflation}},  {\em Astron. Astrophys.} {\bf 594} (2016)
	A20, [\href{http://xxx.lanl.gov/abs/1502.0211}{{\tt arXiv:1502.0211}}].
	
	\bibitem{Ade:2015ava}
	{\bf Planck} Collaboration, P.~Ade {\em et.~al.}, {\it {Planck 2015 results.
			XVII. Constraints on primordial non-Gaussianity}},  {\em Astron. Astrophys.}
	{\bf 594} (2016) A17, [\href{http://xxx.lanl.gov/abs/1502.0159}{{\tt
			arXiv:1502.0159}}].
	
	\bibitem{Planck:2018jri}
	{\bf Planck} Collaboration, Y.~Akrami {\em et.~al.}, {\it {Planck 2018 results.
			X. Constraints on inflation}},  {\em Astron. Astrophys.} {\bf 641} (2020)
	A10, [\href{http://xxx.lanl.gov/abs/1807.0621}{{\tt arXiv:1807.0621}}].
	
	\bibitem{BICEP:2021xfz}
	{\bf BICEP, Keck} Collaboration, P.~A.~R. Ade {\em et.~al.}, {\it {Improved
			Constraints on Primordial Gravitational Waves using Planck, WMAP, and
			BICEP/Keck Observations through the 2018 Observing Season}},  {\em Phys. Rev.
		Lett.} {\bf 127} (2021), no.~15 151301,
	[\href{http://xxx.lanl.gov/abs/2110.0048}{{\tt arXiv:2110.0048}}].
	
	\bibitem{Galloni:2022mok}
	G.~Galloni, N.~Bartolo, S.~Matarrese, M.~Migliaccio, A.~Ricciardone, and
	N.~Vittorio, {\it {Updated constraints on amplitude and tilt of the tensor
			primordial spectrum}},  {\em JCAP} {\bf 04} (2023) 062,
	[\href{http://xxx.lanl.gov/abs/2208.0018}{{\tt arXiv:2208.0018}}].
	
	\bibitem{mclachlan1947theory}
	N.~McLachlan, {\em Theory and Application of Mathieu Functions}.
	\newblock Clarendon Press, 1947.
	
	\bibitem{Dolgov:1982th}
	A.~D. Dolgov and A.~D. Linde, {\it {Baryon Asymmetry in Inflationary
			Universe}},  {\em Phys. Lett. B} {\bf 116} (1982) 329.
	
	\bibitem{Abbott:1982hn}
	L.~F. Abbott, E.~Farhi, and M.~B. Wise, {\it {Particle Production in the New
			Inflationary Cosmology}},  {\em Phys. Lett.} {\bf 117B} (1982) 29.
	
	\bibitem{Traschen:1990sw}
	J.~H. Traschen and R.~H. Brandenberger, {\it {Particle Production During
			Out-of-equilibrium Phase Transitions}},  {\em Phys. Rev. D} {\bf 42} (1990)
	2491--2504.
	
	\bibitem{Kofman:1994rk}
	L.~Kofman, A.~D. Linde, and A.~A. Starobinsky, {\it {Reheating after
			inflation}},  {\em Phys. Rev. Lett.} {\bf 73} (1994) 3195--3198,
	[\href{http://xxx.lanl.gov/abs/hep-th/9405187}{{\tt hep-th/9405187}}].
	
	\bibitem{Shtanov:1994ce}
	Y.~Shtanov, J.~H. Traschen, and R.~H. Brandenberger, {\it {Universe reheating
			after inflation}},  {\em Phys. Rev. D} {\bf 51} (1995) 5438--5455,
	[\href{http://xxx.lanl.gov/abs/hep-ph/9407247}{{\tt hep-ph/9407247}}].
	
	\bibitem{Kofman:1997yn}
	L.~Kofman, A.~D. Linde, and A.~A. Starobinsky, {\it {Towards the theory of
			reheating after inflation}},  {\em Phys. Rev.} {\bf D56} (1997) 3258--3295,
	[\href{http://xxx.lanl.gov/abs/hep-ph/9704452}{{\tt hep-ph/9704452}}].
	
	\bibitem{Liddle:2003as}
	A.~R. Liddle and S.~M. Leach, {\it {How long before the end of inflation were
			observable perturbations produced?}},  {\em Phys. Rev.} {\bf D68} (2003)
	103503, [\href{http://xxx.lanl.gov/abs/astro-ph/0305263}{{\tt
			astro-ph/0305263}}].
	
	\bibitem{Martin:2006rs}
	J.~Martin and C.~Ringeval, {\it {Inflation after WMAP3: Confronting the
			Slow-Roll and Exact Power Spectra to CMB Data}},  {\em JCAP} {\bf 0608}
	(2006) 009, [\href{http://xxx.lanl.gov/abs/astro-ph/0605367}{{\tt
			astro-ph/0605367}}].
	
	\bibitem{Lorenz:2007ze}
	L.~Lorenz, J.~Martin, and C.~Ringeval, {\it {Brane inflation and the WMAP data:
			A Bayesian analysis}},  {\em JCAP} {\bf 04} (2008) 001,
	[\href{http://xxx.lanl.gov/abs/0709.3758}{{\tt arXiv:0709.3758}}].
	
	\bibitem{Korsch2008}
	H.~Korsch and H.~Jodl, {\em Ordinary Differential Equations}.
	\newblock Springer Berlin Heidelberg, Berlin, Heidelberg, 2008.
	
	\bibitem{Allahverdi:2010xz}
	R.~Allahverdi, R.~Brandenberger, F.-Y. Cyr-Racine, and A.~Mazumdar, {\it
		{Reheating in Inflationary Cosmology: Theory and Applications}},  {\em Ann.
		Rev. Nucl. Part. Sci.} {\bf 60} (2010) 27--51,
	[\href{http://xxx.lanl.gov/abs/1001.2600}{{\tt arXiv:1001.2600}}].
	
	\bibitem{Martin:2010kz}
	J.~Martin and C.~Ringeval, {\it {First CMB Constraints on the Inflationary
			Reheating Temperature}},  {\em Phys. Rev.} {\bf D82} (2010) 023511,
	[\href{http://xxx.lanl.gov/abs/1004.5525}{{\tt arXiv:1004.5525}}].
	
	\bibitem{Adshead:2010mc}
	P.~Adshead, R.~Easther, J.~Pritchard, and A.~Loeb, {\it {Inflation and the
			Scale Dependent Spectral Index: Prospects and Strategies}},  {\em JCAP} {\bf
		02} (2011) 021, [\href{http://xxx.lanl.gov/abs/1007.3748}{{\tt
			arXiv:1007.3748}}].
	
	\bibitem{Mielczarek:2010ag}
	J.~Mielczarek, {\it {Reheating temperature from the CMB}},  {\em Phys. Rev. D}
	{\bf 83} (2011) 023502, [\href{http://xxx.lanl.gov/abs/1009.2359}{{\tt
			arXiv:1009.2359}}].
	
	\bibitem{Easther:2011yq}
	R.~Easther and H.~V. Peiris, {\it {Bayesian Analysis of Inflation II: Model
			Selection and Constraints on Reheating}},  {\em Phys. Rev. D} {\bf 85} (2012)
	103533, [\href{http://xxx.lanl.gov/abs/1112.0326}{{\tt arXiv:1112.0326}}].
	
	\bibitem{Choudhury:2013qza}
	S.~Choudhury and A.~Dasgupta, {\it {Galileogenesis: A new cosmophenomenological
			zip code for reheating through R-parity violating coupling}},  {\em Nucl.
		Phys. B} {\bf 882} (2014) 195--204,
	[\href{http://xxx.lanl.gov/abs/1309.1934}{{\tt arXiv:1309.1934}}].
	
	\bibitem{Amin:2014eta}
	M.~A. Amin, M.~P. Hertzberg, D.~I. Kaiser, and J.~Karouby, {\it
		{Nonperturbative Dynamics Of Reheating After Inflation: A Review}},  {\em
		Int. J. Mod. Phys.} {\bf D24} (2014) 1530003,
	[\href{http://xxx.lanl.gov/abs/1410.3808}{{\tt arXiv:1410.3808}}].
	
	\bibitem{Dai:2014jja}
	L.~Dai, M.~Kamionkowski, and J.~Wang, {\it {Reheating constraints to
			inflationary models}},  {\em Phys. Rev. Lett.} {\bf 113} (2014) 041302,
	[\href{http://xxx.lanl.gov/abs/1404.6704}{{\tt arXiv:1404.6704}}].
	
	\bibitem{Martin:2014nya}
	J.~Martin, C.~Ringeval, and V.~Vennin, {\it {Observing Inflationary
			Reheating}},  {\em Phys. Rev. Lett.} {\bf 114} (2015), no.~8 081303,
	[\href{http://xxx.lanl.gov/abs/1410.7958}{{\tt arXiv:1410.7958}}].
	
	\bibitem{Domcke:2015iaa}
	V.~Domcke and J.~Heisig, {\it {Constraints on the reheating temperature from
			sizable tensor modes}},  {\em Phys. Rev. D} {\bf 92} (2015), no.~10 103515,
	[\href{http://xxx.lanl.gov/abs/1504.0034}{{\tt arXiv:1504.0034}}].
	
	\bibitem{Maity:2016uyn}
	D.~Maity and P.~Saha, {\it {Minimal inflationary cosmologies and constraints on
			reheating}},  \href{http://xxx.lanl.gov/abs/1610.0017}{{\tt
			arXiv:1610.0017}}.
	
	\bibitem{Lozanov:2016hid}
	K.~D. Lozanov and M.~A. Amin, {\it {Equation of State and Duration to Radiation
			Domination after Inflation}},  {\em Phys. Rev. Lett.} {\bf 119} (2017), no.~6
	061301, [\href{http://xxx.lanl.gov/abs/1608.0121}{{\tt arXiv:1608.0121}}].
	
	\bibitem{Maity:2018qhi}
	D.~Maity and P.~Saha, {\it {(P)reheating after minimal Plateau Inflation and
			constraints from CMB}},  \href{http://xxx.lanl.gov/abs/1811.1117}{{\tt
			arXiv:1811.1117}}.
	
	\bibitem{Kabir:2016kdh}
	R.~Kabir, A.~Mukherjee, and D.~Lohiya, {\it {Reheating constraints on K\"ahler
			moduli inflation}},  {\em Mod. Phys. Lett. A} {\bf 34} (2019), no.~15
	1950114, [\href{http://xxx.lanl.gov/abs/1609.0924}{{\tt arXiv:1609.0924}}].
	
	\bibitem{Maity:2019ltu}
	D.~Maity and P.~Saha, {\it {Minimal plateau inflationary cosmologies and
			constraints from reheating}},  {\em Class. Quant. Grav.} {\bf 36} (2019)
	045010, [\href{http://xxx.lanl.gov/abs/1902.0189}{{\tt arXiv:1902.0189}}].
	
	\bibitem{Odintsov:2023lbb}
	S.~D. Odintsov and T.~Paul, {\it {From inflation to reheating and their
			dynamical stability analysis in Gauss\textendash{}Bonnet gravity}},  {\em
		Phys. Dark Univ.} {\bf 42} (2023) 101263,
	[\href{http://xxx.lanl.gov/abs/2305.1911}{{\tt arXiv:2305.1911}}].
	
	\bibitem{Chowdhury:2023jft}
	D.~Chowdhury and A.~Hait, {\it {Thermalization in the presence of a
			time-dependent dissipation and its impact on dark matter production}},  {\em
		JHEP} {\bf 09} (2023) 085, [\href{http://xxx.lanl.gov/abs/2302.0665}{{\tt
			arXiv:2302.0665}}].
	
	\bibitem{ElBourakadi:2021blc}
	K.~El~Bourakadi, {\it {Preheating and Reheating after Standard Inflation}},
	\href{http://xxx.lanl.gov/abs/2104.1055}{{\tt arXiv:2104.1055}}.
	
	\bibitem{Nandi:2019xve}
	D.~Nandi and P.~Saha, {\it {Einstein or Jordan: seeking answers from the
			reheating constraints}},  \href{http://xxx.lanl.gov/abs/1907.1029}{{\tt
			arXiv:1907.1029}}.
	
	\bibitem{Turner:1983he}
	M.~S. Turner, {\it {Coherent Scalar Field Oscillations in an Expanding
			Universe}},  {\em Phys. Rev.} {\bf D28} (1983) 1243.
	
	\bibitem{Martin:2013tda}
	J.~Martin, C.~Ringeval, and V.~Vennin, {\it {Encyclopædia Inflationaris}},
	{\em Phys. Dark Univ.} {\bf 5-6} (2014) 75--235,
	[\href{http://xxx.lanl.gov/abs/1303.3787}{{\tt arXiv:1303.3787}}].
	
	\bibitem{PRISM:2013ybg}
	{\bf PRISM} Collaboration, P.~Andre {\em et.~al.}, {\it {PRISM (Polarized
			Radiation Imaging and Spectroscopy Mission): A White Paper on the Ultimate
			Polarimetric Spectro-Imaging of the Microwave and Far-Infrared Sky}},
	\href{http://xxx.lanl.gov/abs/1306.2259}{{\tt arXiv:1306.2259}}.
	
	\bibitem{EuclidTheoryWorkingGroup:2012gxx}
	{\bf Euclid Theory Working Group} Collaboration, L.~Amendola {\em et.~al.},
	{\it {Cosmology and fundamental physics with the Euclid satellite}},  {\em
		Living Rev. Rel.} {\bf 16} (2013) 6,
	[\href{http://xxx.lanl.gov/abs/1206.1225}{{\tt arXiv:1206.1225}}].
	
	\bibitem{Mao:2008ug}
	Y.~Mao, M.~Tegmark, M.~McQuinn, M.~Zaldarriaga, and O.~Zahn, {\it {How
			accurately can 21 cm tomography constrain cosmology?}},  {\em Phys. Rev.}
	{\bf D78} (2008) 023529, [\href{http://xxx.lanl.gov/abs/0802.1710}{{\tt
			arXiv:0802.1710}}].
	
	\bibitem{CORE:2016ymi}
	{\bf CORE} Collaboration, F.~Finelli {\em et.~al.}, {\it {Exploring cosmic
			origins with CORE: Inflation}},  {\em JCAP} {\bf 04} (2018) 016,
	[\href{http://xxx.lanl.gov/abs/1612.0827}{{\tt arXiv:1612.0827}}].
	
	\bibitem{Jedamzik:2010dq}
	K.~Jedamzik, M.~Lemoine, and J.~Martin, {\it {Collapse of Small-Scale Density
			Perturbations during Preheating in Single Field Inflation}},  {\em JCAP} {\bf
		09} (2010) 034, [\href{http://xxx.lanl.gov/abs/1002.3039}{{\tt
			arXiv:1002.3039}}].
	
	\bibitem{Sharma:2018kgs}
	R.~Sharma, K.~Subramanian, and T.~R. Seshadri, {\it {Generation of helical
			magnetic field in a viable scenario of inflationary magnetogenesis}},  {\em
		Phys. Rev. D} {\bf 97} (2018), no.~8 083503,
	[\href{http://xxx.lanl.gov/abs/1802.0484}{{\tt arXiv:1802.0484}}].
	
	\bibitem{Martin:2019nuw}
	J.~Martin, T.~Papanikolaou, and V.~Vennin, {\it {Primordial black holes from
			the preheating instability in single-field inflation}},  {\em JCAP} {\bf 01}
	(2020) 024, [\href{http://xxx.lanl.gov/abs/1907.0423}{{\tt
			arXiv:1907.0423}}].
	
	\bibitem{Auclair:2020csm}
	P.~Auclair and V.~Vennin, {\it {Primordial black holes from metric preheating:
			mass fraction in the excursion-set approach}},  {\em JCAP} {\bf 02} (2021)
	038, [\href{http://xxx.lanl.gov/abs/2011.0563}{{\tt arXiv:2011.0563}}].
	
	\bibitem{Haque:2020bip}
	M.~R. Haque, D.~Maity, and S.~Pal, {\it {Probing the reheating phase through
			primordial magnetic field and CMB}},  {\em Phys. Rev. D} {\bf 103} (2021),
	no.~10 103540, [\href{http://xxx.lanl.gov/abs/2012.1085}{{\tt
			arXiv:2012.1085}}].
	
	\bibitem{Bamba:2021wyx}
	K.~Bamba, S.~D. Odintsov, T.~Paul, and D.~Maity, {\it {Helical magnetogenesis
			with reheating phase from higher curvature coupling and baryogenesis}},  {\em
		Phys. Dark Univ.} {\bf 36} (2022) 101025,
	[\href{http://xxx.lanl.gov/abs/2107.1152}{{\tt arXiv:2107.1152}}].
	
	\bibitem{Lozanov:2019jxc}
	K.~D. Lozanov, {\it {Lectures on Reheating after Inflation}},
	\href{http://xxx.lanl.gov/abs/1907.0440}{{\tt arXiv:1907.0440}}.
	
	\bibitem{Kwapisz:2019cxq}
	J.~H. Kwapisz, {\it {Conformal standard model and inflation}},
	\href{http://xxx.lanl.gov/abs/1911.0477}{{\tt arXiv:1911.0477}}.
	
	\bibitem{Mukhanov:2005sc}
	V.~Mukhanov, {\em {Physical Foundations of Cosmology}}.
	\newblock Cambridge University Press, Oxford, 2005.
	
	\bibitem{Urena-Lopez:2016yon}
	L.~A. Ure\~na L\'{o}pez, {\it {Scalar fields in Cosmology: dark matter and
			inflation}},  {\em J. Phys. Conf. Ser.} {\bf 761} (2016), no.~1 012076.
	
	\bibitem{Kallosh:2013yoa}
	R.~Kallosh, A.~Linde, and D.~Roest, {\it {Superconformal Inflationary
			$\alpha$-Attractors}},  {\em JHEP} {\bf 11} (2013) 198,
	[\href{http://xxx.lanl.gov/abs/1311.0472}{{\tt arXiv:1311.0472}}].
	
	\bibitem{Kallosh:2013tua}
	R.~Kallosh, A.~Linde, and D.~Roest, {\it {Universal Attractor for Inflation at
			Strong Coupling}},  {\em Phys. Rev. Lett.} {\bf 112} (2014), no.~1 011303,
	[\href{http://xxx.lanl.gov/abs/1310.3950}{{\tt arXiv:1310.3950}}].
	
	\bibitem{Galante:2014ifa}
	M.~Galante, R.~Kallosh, A.~Linde, and D.~Roest, {\it {Unity of Cosmological
			Inflation Attractors}},  {\em Phys. Rev. Lett.} {\bf 114} (2015), no.~14
	141302, [\href{http://xxx.lanl.gov/abs/1412.3797}{{\tt arXiv:1412.3797}}].
	
	\bibitem{Carrasco:2015pla}
	J.~J.~M. Carrasco, R.~Kallosh, and A.~Linde, {\it {$\alpha $-Attractors:
			Planck, LHC and Dark Energy}},  {\em JHEP} {\bf 10} (2015) 147,
	[\href{http://xxx.lanl.gov/abs/1506.0170}{{\tt arXiv:1506.0170}}].
	
	\bibitem{Kallosh:2022feu}
	R.~Kallosh and A.~Linde, {\it {Polynomial \ensuremath{\alpha}-attractors}},
	{\em JCAP} {\bf 04} (2022), no.~04 017,
	[\href{http://xxx.lanl.gov/abs/2202.0649}{{\tt arXiv:2202.0649}}].
	
	\bibitem{Bhattacharya:2022akq}
	S.~Bhattacharya, K.~Dutta, M.~R. Gangopadhyay, and A.~Maharana, {\it
		{\ensuremath{\alpha}-attractor inflation: Models and predictions}},  {\em
		Phys. Rev. D} {\bf 107} (2023), no.~10 103530,
	[\href{http://xxx.lanl.gov/abs/2212.1336}{{\tt arXiv:2212.1336}}].
	
	\bibitem{Boyle:2007zx}
	L.~A. Boyle and A.~Buonanno, {\it {Relating gravitational wave constraints from
			primordial nucleosynthesis, pulsar timing, laser interferometers, and the
			CMB: Implications for the early Universe}},  {\em Phys. Rev.} {\bf D78}
	(2008) 043531, [\href{http://xxx.lanl.gov/abs/0708.2279}{{\tt
			arXiv:0708.2279}}].
	
	\bibitem{Koh:2018qcy}
	S.~Koh, B.-H. Lee, and G.~Tumurtushaa, {\it {Constraints on the reheating
			parameters after Gauss-Bonnet inflation from primordial gravitational
			waves}},  {\em Phys. Rev.} {\bf D98} (2018), no.~10 103511,
	[\href{http://xxx.lanl.gov/abs/1807.0442}{{\tt arXiv:1807.0442}}].
	
	\bibitem{Garcia-Bellido:2017mdw}
	J.~Garcia-Bellido and E.~Ruiz~Morales, {\it {Primordial black holes from single
			field models of inflation}},  {\em Phys. Dark Univ.} {\bf 18} (2017) 47--54,
	[\href{http://xxx.lanl.gov/abs/1702.0390}{{\tt arXiv:1702.0390}}].
	
	\bibitem{Germani:2017bcs}
	C.~Germani and T.~Prokopec, {\it {On primordial black holes from an inflection
			point}},  {\em Phys. Dark Univ.} {\bf 18} (2017) 6--10,
	[\href{http://xxx.lanl.gov/abs/1706.0422}{{\tt arXiv:1706.0422}}].
	
	\bibitem{Ballesteros:2017fsr}
	G.~Ballesteros and M.~Taoso, {\it {Primordial black hole dark matter from
			single field inflation}},  {\em Phys. Rev. D} {\bf 97} (2018), no.~2 023501,
	[\href{http://xxx.lanl.gov/abs/1709.0556}{{\tt arXiv:1709.0556}}].
	
	\bibitem{Dalianis:2018frf}
	I.~Dalianis, A.~Kehagias, and G.~Tringas, {\it {Primordial black holes from
			$\alpha$-attractors}},  {\em JCAP} {\bf 01} (2019) 037,
	[\href{http://xxx.lanl.gov/abs/1805.0948}{{\tt arXiv:1805.0948}}].
	
	\bibitem{Maldacena2003}
	J.~M. Maldacena, {\it {Non-Gaussian features of primordial fluctuations in
			single field inflationary models}},  {\em JHEP} {\bf 0305} (2003) 013,
	[\href{http://xxx.lanl.gov/abs/astro-ph/0210603}{{\tt astro-ph/0210603}}].
	
	\bibitem{Chen:2006nt}
	X.~Chen, M.-x. Huang, S.~Kachru, and G.~Shiu, {\it {Observational signatures
			and non-Gaussianities of general single field inflation}},  {\em JCAP} {\bf
		0701} (2007) 002, [\href{http://xxx.lanl.gov/abs/hep-th/0605045}{{\tt
			hep-th/0605045}}].
	
	\bibitem{Nandi:2015ogk}
	D.~Nandi and S.~Shankaranarayanan, {\it {Complete Hamiltonian analysis of
			cosmological perturbations at all orders}},  {\em JCAP} {\bf 1606} (2016),
	no.~06 038, [\href{http://xxx.lanl.gov/abs/1512.0253}{{\tt
			arXiv:1512.0253}}].
	
	\bibitem{Nandi:2016pfr}
	D.~Nandi and S.~Shankaranarayanan, {\it {Complete Hamiltonian analysis of
			cosmological perturbations at all orders II: Non-canonical scalar field}},
	{\em JCAP} {\bf 1610} (2016) 008,
	[\href{http://xxx.lanl.gov/abs/1606.0574}{{\tt arXiv:1606.0574}}].
	
	\bibitem{Nandi:2017pfw}
	D.~Nandi, {\it {Hamiltonian formalism of cosmological perturbations and higher
			derivative theories}},  \href{http://xxx.lanl.gov/abs/1707.0297}{{\tt
			arXiv:1707.0297}}.
	
	\bibitem{Nandi:2018ooh}
	D.~Nandi, {\it {Stable contraction in Brans-Dicke cosmology}},  {\em JCAP} {\bf
		1905} (2018) 040, [\href{http://xxx.lanl.gov/abs/1811.0962}{{\tt
			arXiv:1811.0962}}].
	
	\bibitem{Nandi:2019xlj}
	D.~Nandi, {\it {Note on stability in conformally connected frames}},  {\em
		Phys. Rev.} {\bf D99} (2019), no.~10 103532,
	[\href{http://xxx.lanl.gov/abs/1904.0015}{{\tt arXiv:1904.0015}}].
	
	\bibitem{Nandi:2019xag}
	D.~Nandi and L.~Sriramkumar, {\it {Can a nonminimal coupling restore the
			consistency condition in bouncing universes?}},  {\em Phys. Rev.} {\bf D101}
	(2020), no.~4 043506, [\href{http://xxx.lanl.gov/abs/1904.1325}{{\tt
			arXiv:1904.1325}}].
	
	\bibitem{Nandi:2020sif}
	D.~Nandi, {\it {Bounce from Inflation}},  {\em Phys. Lett. B} {\bf 809} (2020)
	135695, [\href{http://xxx.lanl.gov/abs/2003.0206}{{\tt arXiv:2003.0206}}].
	
	\bibitem{Nandi:2020szp}
	D.~Nandi, {\it {Stability of a viable non-minimal bounce}},  {\em Universe}
	{\bf 7} (2021), no.~3 62, [\href{http://xxx.lanl.gov/abs/2009.0313}{{\tt
			arXiv:2009.0313}}].
	
	\bibitem{Nandi:2022twa}
	D.~Nandi and M.~Kaur, {\it {Viable bounce from non-minimal inflation}},
	\href{http://xxx.lanl.gov/abs/2206.0833}{{\tt arXiv:2206.0833}}.
	
	\bibitem{Nandi:2023ooo}
	D.~Nandi and M.~Kaur, {\it {Inflation vs. Ekpyrosis -- comparing stability in
			general non-minimal theory}},  \href{http://xxx.lanl.gov/abs/2302.0341}{{\tt
			arXiv:2302.0341}}.
	
	\bibitem{Kaur:2023uaz}
	M.~Kaur, D.~Nandi, D.~Choudhury, and T.~R. Seshadri, {\it {Universe bouncing
			its way to inflation}},  \href{http://xxx.lanl.gov/abs/2302.1369}{{\tt
			arXiv:2302.1369}}.
	
\end{thebibliography}
\providecommand{\href}[2]{#2}\begingroup\raggedright\endgroup
\end{document}